\documentclass[prl,twocolumn,superscriptaddress,longbibliography]{revtex4-2}
\usepackage{graphicx}
\usepackage{dcolumn}
\usepackage{bm}

\usepackage{amsmath,amssymb,bm}
\usepackage{wasysym}
\usepackage{graphicx}
\usepackage{epstopdf}
\usepackage{latexsym}
\usepackage{subfigure}
\usepackage[usenames, dvipsnames]{color}
\usepackage[usenames, dvipsnames]{xcolor}
\usepackage{natbib}
\usepackage{braket}
\usepackage{float}
\usepackage[normalem]{ulem}
\usepackage{comment}
\usepackage{mathtools}
\usepackage{array}
\usepackage{tabu}
\usepackage{multirow}
\usepackage{chemformula}
\usepackage{svg}
\usepackage[T1]{fontenc}
\usepackage[	
citecolor=ForestGreen,
colorlinks, 
linkcolor=violet,
urlcolor=black,
]{hyperref}
\newcommand\redsout{\bgroup\markoverwith{\textcolor{red}{\rule[0.5ex]{2pt}{0.4pt}}}\ULon}
\newcommand\bluesout{\bgroup\markoverwith{\textcolor{black}{\rule[0.5ex]{2pt}{0.4pt}}}\ULon}

\newcommand{\prlsec}[1]{{{\it #1:--}}}

\newcommand{\SPhide}[1]{{}}

\newcommand{\SPEDITOKAY}[2]{{}{\textcolor{black}{#2}}}

\def\acz2{ac-$\mathbb{Z}_2$}

\begin{document}

\title{Eigenstate thermalization at the edge of the many-body spectrum}

\author{Arnav Pushkar}
\email{arnavpushkar@gmail.com}
\affiliation{Department of Physics, Indian Institute of Technology Bombay, Mumbai 400076, India.}
\affiliation{Department of Physics, Kyoto University, Kitashirakawa Oiwakecho, Sakyo-ku, Kyoto 606-8502, Japan.}
\altaffiliation{On visit to Department of Physics, Kyoto University, Kyoto 606-8502, Japan}
\author{Pranay Patil}
\affiliation{Department of Physics, Indian Institute of Technology Madras, Chennai 600036, India.}
\affiliation{Theory of Quantum Matter Unit, Okinawa Institute of Science and Technology Graduate University, Onna-son, Okinawa 904-0412, Japan}
\author{Soumya Bera}
\affiliation{Department of Physics, Indian Institute of Technology Bombay, Mumbai 400076, India.}
\author{Masaki Tezuka}
\affiliation{Department of Physics, Kyoto University, Kitashirakawa Oiwakecho, Sakyo-ku, Kyoto 606-8502, Japan.}
\author{Sumiran Pujari}
\email{sumiran.pujari@iitb.ac.in}
\affiliation{Department of Physics, Indian Institute of Technology Bombay, Mumbai 400076, India.}

\date{\today}

\begin{abstract}
It is well-known that thermalization 
breaks down in presence of an extensive number of symmetries or conservation laws that are mutually compatible or commuting with each other.  
We ask here how an extensive number of mutually incompatible local symmetries govern eigenstate thermalization (ETH).
We will investigate this through models within the class of bond-dependent $\mathbb{Z}_2$-symmetric quantum spin-$\frac{1}{2}$ Hamiltonians that naturally admit this structure.
This leads to exponentially large degeneracies in the spectrum
governed by the incompatible symmetry structure.
Here, using spectral diagnostics, we show that non-integrability can survive despite extensively many conserved quantities when they are incompatible.
Furthermore, we show that ETH behavior persists deep into the many-body spectral edge due to this structure.
In other words, incompatible local symmetries enable a mechanism to obtain enough thermodynamic entropy at low excitation energy densities to make ETH operational or predictive near the ground state.
\end{abstract}
\maketitle

The eigenstate thermalization hypothesis (ETH) is by now the established framework for understanding how isolated quantum many-body systems approach thermal equilibrium~\cite{Deutsch1991,Srednicki,RigolNature,DAlessio2016,Deutsch2018,Mori2018}. 
A seminal cold atoms study is Ref.~\cite{Kinoshita_2006} from 2006 where breakdown of ETH was also observed in one dimension.  
It continues to receive significant attention due to new experiments enabled by rapid, ongoing progress in quantum simulation or emulation of isolated quantum systems.
Such breakdown is often due to the presence of extensively many conserved quantities or ``conservation laws'' -- for instance in integrable models -- 
and require a generalized Gibbs ensemble description~\cite{Rigol_etal_2007,Rigol_Muramatsu_Olshanii_2006,Rigol_2009} instead of the standard Gibbs ensemble. 
These conserved quantities being extensive in number drastically reduce the Hilbert space size that the system can explore under unitary dynamics thus breaking ergodicity.
This reduction happens if the conserved quantities are mutually compatible or commuting.
Thus in situations where there are extensive number of conserved quantities yet not all mutually compatible, what is the fate of ETH? Does it survive or break down?

We will investigate the above using a set of models from the class of bond-dependent $\mathbb{Z}_2$ quantum spin Hamiltonians also referred to as compass models~\cite{compass_models_review}. 
A different question which is not the focus here concerns the fate of ETH in presence of a few $\mathrm{O}(1)$ incompatible conserved quantities that has received some attention and termed as non-Abelian ETH~\cite{Murthy_etal_nonabETH_2023,nonabETH_review,Lasek_Hapern_2026}.
The $\mathbb{Z}_2$ model class also contains for example the well-known Kitaev honeycomb model. 
It has commuting or compatible local plaquette $\mathbb{Z}_2$ charges that in fact render it integrable. 
The models under study here naturally admit an extensive number of mutually incompatible  conserved quantities (MICQ) that are also local $\mathbb{Z}_2$ charges.
This gives us a model setting to investigate the general question posed above.

We first show that non-integrability or ergodicity can survive due to incompatibility in presence of an extensive number of conserved quantities.
There is another consequence of extensively many mutually incompatible (local) conserved quantities (EMICQ) that is quite pertinent.
The ETH ansatz requires an extensive thermodynamic entropy $S(\epsilon)$ or equivalently an exponential (many-body) density of states since $S(\epsilon) \equiv k_{\text{B}} \ln \Omega(\epsilon,d\epsilon)$ where $\Omega(\epsilon,d\epsilon)$ counts the number of eigenstates in a tiny energy window $d\epsilon$ around energy $\epsilon$. 
This is almost always present at mid-spectrum \SPEDITOKAY{but generically absent at the spectral edges}{} for local Hamiltonians.
It turns out the above non-commuting local symmetry structure necessarily leads to exponentially macroscopic degeneracies \emph{across the spectrum} that further ensures liquid-like correlations in the ground state~\cite{SP2024}.  
Thus this also has a potential to extend the applicability of ETH deep into the spectral edges. 
We next show that this indeed is possible which is the central result of this work.

\begin{figure}[]
\includegraphics[width=\linewidth]{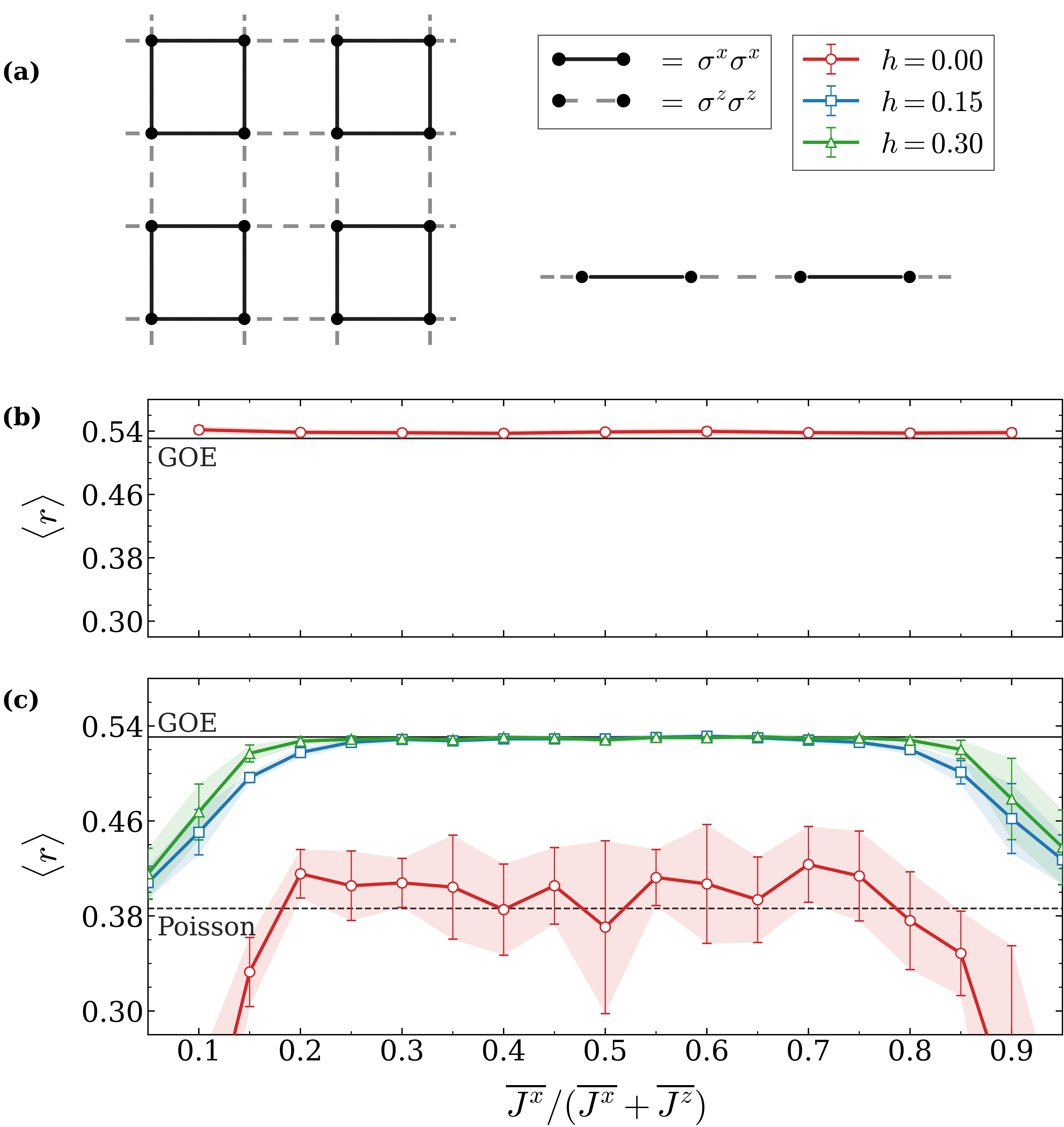}
\caption{\label{fig:1dchain_level_stats} 
(a) Diagrams illustrating the models of Eq.~\ref{eq:2dchecker} and Eq.~\ref{eq:1dchain} with extensively many mutually incompatible local symmetries (abbreviated to EMICQ in main text).
(b) Average gap ratio $\langle r\rangle$ for the 16-site checkerboard model (Eq.~\ref{eq:2dchecker}) with bond disorder over 30 realizations within a randomly chosen (anticommuting) symmetry sector. 
The Gaussian orthogonal ensemble (GOE) value is observed over the full range of $\overline{J^x}/\overline{J^z}$ indicating robust non-integrability.
(c) $\langle r\rangle$ for Eq.~\ref{eq:H1d_with_field} averaged over fifteen disordered realizations and shown vs. $\frac{\overline{J^x}}{\overline{J^x}+\overline{J^z}}$ for the $L=12$ chain.
Bond disorder removes lattice symmetry based degeneracies.
The disorder strength is such that $\Delta J_\mu/\overline{J_\mu} \sim 0.1$ with $\mu \in \{x,z\}$ and the mean value of couplings $\overline{J^x}+\overline{J^z}$ set to 1. 
GOE and Poisson values of $\sim 0.531$ and $\sim 0.386$ are indicated in the plots for reference.
}
\end{figure}

\begin{figure}
\begin{center}
\includegraphics[width=0.99\linewidth]{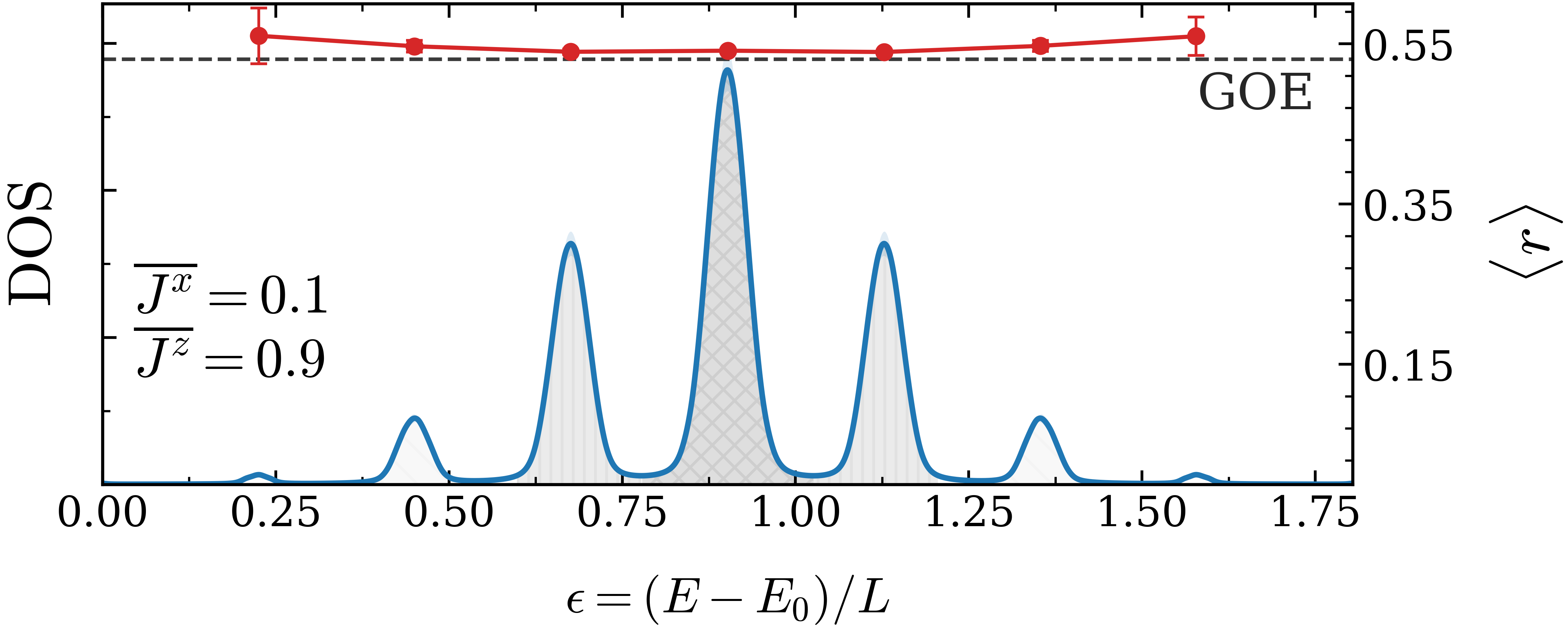}
\caption{\label{fig:2dsquare_level_stats} 
In the limit when one set of checkerboard couplings dominates the other in Eq.~\ref{eq:2dchecker}, the many-body density of states splits into lobes as shown. 
Also shown is the lobe-resolved $\langle r\rangle$ for a $L=16$ system within a randomly chosen (anticommuting) symmetry sector.
GOE statistics is seen in each lobe down to the lowest lobes.
}
\end{center}
\end{figure}

Consider the following $2d$ square lattice model,
\begin{equation}
H_{2d}=
\sum_{\boxed{x}}\Bigg(\sum_{\langle i,j\rangle\in\boxed{x}}J^x_{ij}\sigma^x_i\sigma^x_j\Bigg)
+
\sum_{\boxed{z}}\Bigg(\sum_{\langle i,j\rangle\in\boxed{z}}J^z_{ij}\sigma^z_i\sigma^z_j\Bigg).
\label{eq:2dchecker}
\end{equation}
from the general class of EMICQ models~\cite{SP2024,PujariNigam2025,acqsl_name_note}.
$\boxed{x}$, $\boxed{z}$ represent square plaquettes marked as ``$x$'' or ``$z$'' that are arranged alternately in a checkerboard fashion (Fig.~\ref{fig:2dsquare_level_stats}(a).
$\sigma^\mu_i$ with $\mu \in \{x,y,z\}$ are the standard Pauli matrices for spin-$\frac{1}{2}$ degrees of freedom at site $i$.
The incompatible symmetries here are $\prod_{i \in \boxed{x}}\sigma^z_i$ for $\boxed{x}$ plaquettes and $\prod_{i \in \boxed{z}}\sigma^x_i$ for $\boxed{z}$ plaquettes. 
They anticommute when they share a site which leads to extensive degeneracies~\cite{SP2024}.
Fig.~\ref{fig:1dchain_level_stats}(a) shows the average gap ratio~\cite{OganesyanHuse2007} of the energy level spacings within each symmetry sector for different parameter values of Eq.~\ref{eq:2dchecker}.
It tends to the Gaussian orthogonal ensemble (GOE) value for all parameter values showing that $H_{2d}$ is non-integrable. 
The level spacing histograms also show GOE behavior~\cite{supp_info, Atas2013}.

Let us also consider a related $1d$ chain model that will allow us to study finite size effects over a wider range of sizes.
It is
\begin{equation}
H_{1d}
= \sum_{\langle i,j\rangle_x} J^x_{ij}\,\sigma^x_i \sigma^x_j
+ \sum_{\langle i,j\rangle_z} J^z_{ij}\,\sigma^z_i \sigma^z_j.
\label{eq:1dchain}
\end{equation}
where $\langle i,j\rangle_x$ and $\langle i,j\rangle_z$ are sets of bonds marked as ``$x$'' or ``$z$'' that are arranged alternately as shown in Fig.~\ref{fig:1dchain_level_stats}(b).
Such one-dimensional bond-dependent Ising interactions were notably found to be relevant to the phenomenology of the chain compound Cobalt Niobate (CoNb$_2$O$_6$)~\cite{Morris_Desai_etal_natphys_2021}.
The EMICQs for Eq.~\ref{eq:1dchain} are $\sigma^z_i \sigma^z_j$ on $\{\langle i,j\rangle_x\}$ bond set and $\sigma^x_i \sigma^x_j$ on $\{\langle i,j\rangle_z\}$ bond set.
This model can also be thought of as a stripped chain version of the Kitaev honeycomb model.
Thus it is also free fermionizable and hence integrable. 
This is seen in the behavior of the average gap ratio $\langle r\rangle$
over the full range of $J^x/J^z$ for $h=0$ in Fig.~\ref{fig:1dchain_level_stats}(a)~\cite{poisson_fragmentation_note}. 
To break integrability, we add a canted field in the $\left(\frac{1}{\sqrt{2}},0,\frac{1}{\sqrt{2}}\right)$ direction, i.e. 
\begin{equation}
    H_{1d} + h \sum_i (\sigma^x_i+\sigma^z_i)/\sqrt{2}
    \label{eq:H1d_with_field}
\end{equation}
The average gap ratio for Eq.~\ref{eq:H1d_with_field} for $h=\{0.15,0.30\}$ is shown in Fig.~\ref{fig:1dchain_level_stats} where we see non-integrable behavior.
The canted field voids free fermionization leading to level repulsion generically across the spectrum.
GOE statistics is also present for other field directions except for $x$ and $z$ for which integrability remains.
For extreme ends in Fig.~\ref{fig:1dchain_level_stats}(b), we see deviations from GOE at these system sizes.
Analogous routes to non-integrability in spin chains have been studied before~\cite{Santos2004,SantosRigol2010,OganesyanHuse2007}.
The canted field also voids the local conservation laws.
However, this does not void the physics we are interested in. 
Rather, we think of it as a suitable perturbation that breaks integrability of Eq.~\ref{eq:1dchain}.
We will show in what follows that eigenstate thermalization extends down to the many-body spectral edge for such non-integrable EMICQ models.
This ``spectral edge ETH effect'' thus needs the canted field on top of the EMICQ structure due the lack of non-integrability for Eq.~\ref{eq:1dchain} (Eq.~\ref{eq:H1d_with_field} with $h=0$).
As we have seen in Fig.~\ref{fig:1dchain_level_stats}(a), the $2d$ checkerboard model is non-integrable without any applied field and thus this effect would be ``intrinsically'' present for generic tiny perturbations, e.g. a tiny amount of field noise from environment.

From the above discussion, we see that the two-dimensional model of Eq.~\ref{eq:2dchecker} is a square lattice extension of Eq.~\ref{eq:1dchain},
and is thus a distinct non-integrable higher-dimensional extension of the integrable $1d$ Kitaev-type chain model in Eq.~\ref{eq:1dchain} when compared to the integrable Kitaev honeycomb model. 
The $\mathbb{Z}_2$ plaquette charges of the honeycomb model remarkably leads to free fermionization~\cite{Kitaev_2006} that renders it integrable, while Eq.~\ref{eq:2dchecker} remains non-integrable in contrast (Fig.~\ref{fig:1dchain_level_stats}(a)) despite extensively many conserved plaquette charges. 
The equilibrium phase diagrams of the translational invariant cases of Eq.~\ref{eq:2dchecker},~\ref{eq:1dchain} assuming thermalization were studied in Refs.~\cite{Wenzel_Janke_2009,Agrapidis_etal_2018} earlier motivated by compass physics~\cite{compass_models_review}. 
Our results here apply to the broader EMICQ model class~\cite{SP2024} -- e.g. imagine a variant of Eq.~\ref{eq:2dchecker} without restricting to nearest neighbors, or on $2d$ kagome or $3d$ pyrochlore lattices with triangular motifs~\cite{PujariNigam2025} including disordered couplings $J^\mu_{ij}$ without loss of generality.

\begin{figure}[]
\centering
\includegraphics[width=\linewidth]{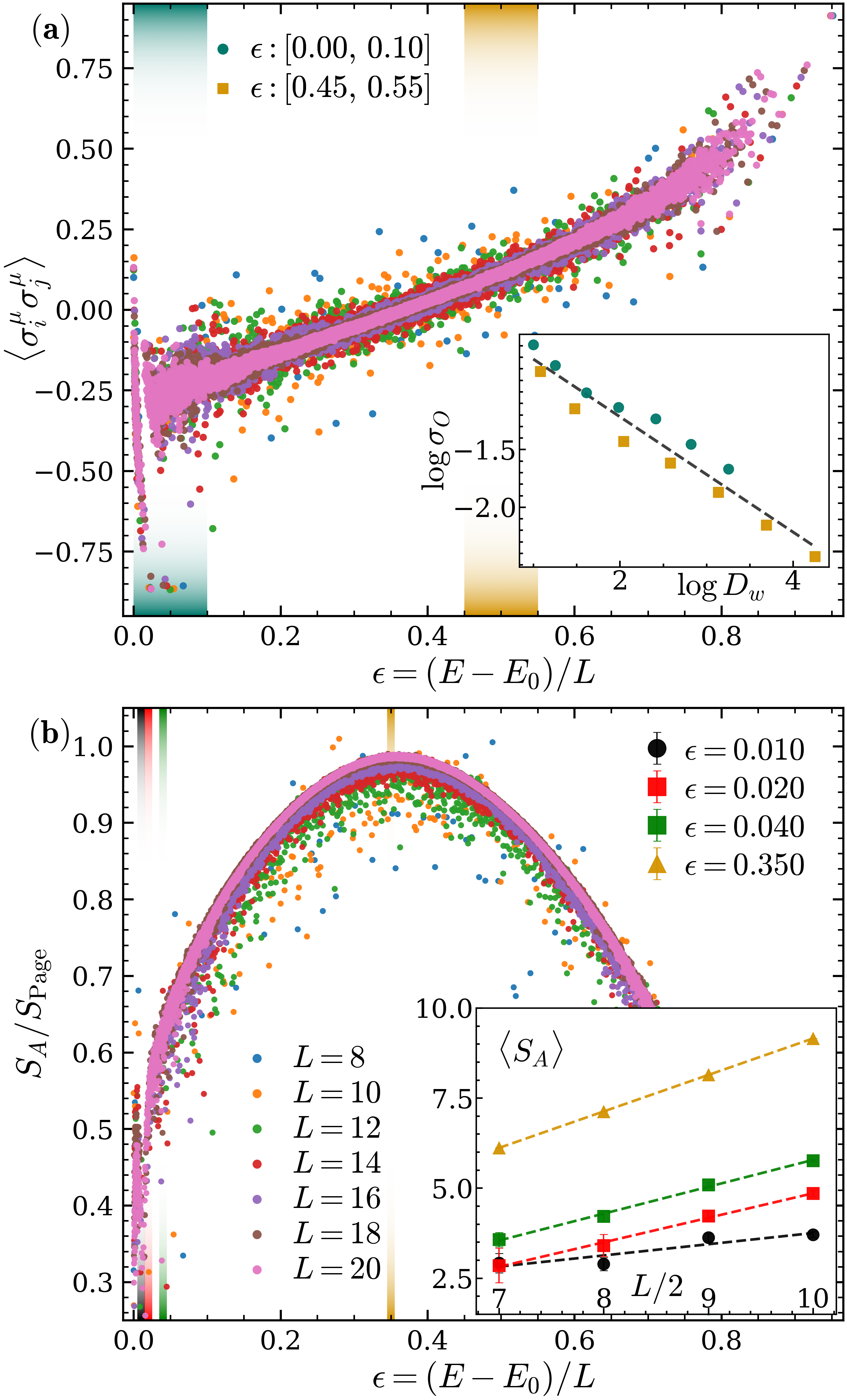}
\caption{\label{fig:1dchain_speckle_page}
(a) Distribution of local observable matrix elements $\langle E|\hat{O}|E\rangle$ versus (excitation) energy density $\epsilon = (E - E_0)/L$ for Eq.~\ref{eq:H1d_with_field} for $h=0.3$ (marked by white triangle in Fig.~\ref{fig:1dchain_level_stats}).
$\hat{O}=\sigma^\mu_i \sigma^\mu_j$ for two randomly chosen sites above with $\mu$ chosen randomly in spin space. 
Here we use translational symmetry ($k=0$ sector) to access bigger system sizes.
Inset: scaling of their fluctuations vs. $D_w$ in log-log in the spectral bulk and at the spectral edge.
The $\epsilon$ windows used to measure the fluctuations are given in the inset legend.
The expected $\sqrt{D_w}$ behavior is shown for comparison as a dotted line. 
(b) Distribution of half-chain EE normalized to the Page limit $\left( \frac{L}{2} - \frac{1}{2\ln 2}\right)$ vs. $\epsilon$~\cite{Page1993}. 
It clearly shows a smooth arc nature across the spectrum which tightens with increasing system size typical of non-integrability.
The inset shows the scaling of the mean EE in $\epsilon$ windows for three excitation energy densities near the spectral edge centered at $\epsilon = \{0.01,0.02,0.04\}$ also indicated by thin strips in the main panel. 
EE scaling from the spectral bulk is also shown for a window centered at $\epsilon=0.4$ indicated by the thin yellow strip in the main panel.
Note the energy window width used for this analysis is $0.01$ which is much smaller than in panel (a).
Both the spectral edge and bulk show volume law growth $\propto L$.
}
\end{figure}

ETH physics also gets reflected in the eigenstates, in particular through the distribution of matrix elements of local observables.
According to the ETH ansatz, their distribution is a smooth function of energy, with fluctuations in a (microcanonical) energy window massively suppressed in presence of extensive thermodynamic entropy. 
The fluctuation suppression goes as $\propto D_w^{-1/2}$ where $D_w$ is the Hilbert space dimension of this window~\cite{Beugeling_etal_pre_2014,KimIkedaHuse2014}.
In practice, for finite sizes one chooses energy windows with enough states to ensure good statistics while estimating the fluctuations.
We show this for a representative non-integrable point $\left(\frac{J^x}{J^x+J^z},h\right)=(0.6,0.3)$ from Fig.~\ref{fig:1dchain_level_stats}(a) as a function of (excitation) energy density $\epsilon = \frac{E - E_0}{L}$ where E0 is the ground state energy. 
These fluctuations are visibly suppressed as $D_w$ increases as seen in the tightening of the distribution for increasing $L$. 
We find that the fluctuations over the full distribution is indeed of the power-law form $D_w^{m}$ with $m<0$~\cite{supp_info} which tends to zero in the thermodynamic limit, but not quite the expected $D_w^{-1/2}$. 
A closer look at the speckle plots in Fig.~\ref{fig:1dchain_speckle_page}(a) shows a noticeable set of outlying states. 
These states persist over the accessible system sizes ($L\leq20$) which are found to cause this deviation. 
To see this, we focus on the region where the matrix elements cluster most and disregard the outlying points (beyond $(\bar{O}+3\sigma_0)$) and compute the fluctuations within the set of these ``typical'' states. 
With this, the fluctuations show scaling in good agreement with ETH scaling of $D_w^{-1/2}$ as shown in Fig.~\ref{fig:1dchain_speckle_page}(a) inset. 
We have also performed fits to a $D^m_w$ form for several different local observables which confirm these conclusions~\cite{supp_info}.
Notably we see this ETH behavior for energy windows from both center \emph{and} edge of the many-body spectrum.
This is the first piece of evidence for EMICQ induced spectral edge ETH. 
The outlying states may be athermal or atypical scar-like states but they do not seem to show appreciably low entanglement usually seen for scars~\cite{scars_review}.
The ratio of the number of these outlying to thermal states is slightly more than expected from equivalent outliers on a bell curve.
We do not pursue the nature of these states in this work.

We shift our attention now to entanglement content of Eq.~\ref{eq:H1d_with_field}. 
It is known that the presence of extensive thermodynamic entropy apart from operationalizing ETH also leads to volume law entanglement~\cite{Deutsch_2010,Deutsch_Li_Sharma_2013,Dymarsky_Lashkari_Liu_2018}.
Thus we will compare entanglement scaling behavior from the spectral edge and spectral bulk as an alternative probe.
We computed the half-chain von Neumann entanglement entropy (EE) for Eq.~\ref{eq:H1d_with_field} with translational symmetry to access bigger system sizes. 
Fig.~\ref{fig:1dchain_speckle_page}(b) main panel shows the distribution of EE data vs. (excitation) energy density $\epsilon$ scaled by the Page limit for all states in the $k=0$ sector for different system sizes $L$. 
The first thing to note is the smooth arc nature of the distribution which also becomes tighter with system size.
This is a key characteristic of non-integrable or thermalizing many-body spectra~\cite{Haque_McClarty_Khaymovich_pre2022} 
in agreement with earlier conclusions from Fig.~\ref{fig:1dchain_level_stats}. 
The maximum of the arc approaches the Page value representative of the infinite temperature limit. 
This behavior is expectedly volume law as shown for the mean EE in Fig.~\ref{fig:1dchain_speckle_page}(b) inset for a tiny energy window in the spectral bulk.
This behavior is independent of the choice of the window size and also seen in other $k$ sectors~\cite{supp_info}.

We now discuss the scaling of mean EE near the spectral edge as shown in Fig.~\ref{fig:1dchain_speckle_page}(b) inset.
We continue to find volume law in this regime as seen from the linear growth of the mean half-chain EE vs. $L$. 
The rate of growth is however slower than the bulk. 
This is to be expected since the spectral edge is (furthest) away from the infinite temperature limit.
This raises an interesting question regarding the thermodynamic limit:
how does the mean EE scaling depend on the order of limits $L \rightarrow \infty$ and 
$\epsilon \rightarrow 0^+$?
By $\epsilon \rightarrow 0^+$, we mean throughout in this paper an amount of excitation energy that is intensive and $\emph{not}$ extensive in the couplings $\epsilon \ll O(J)$.
Clearly, if we first take $\epsilon \rightarrow 0^+$, i.e. $E = E_0$, then EE scaling will follow area law from the ground state (of the local Hamiltonian Eq.~\ref{eq:H1d_with_field}).
We claim that there is volume law EE scaling instead for the opposite order of limits, i.e. $L \rightarrow \infty$ taken first, in presence of an EMICQ structure. 
Evidence for this was already seen in Fig.~\ref{fig:1dchain_speckle_page}(b) inset. 
The above taken together forms strong evidence for the spectral edge ETH effect in the $1d$ model of Eq.~\ref{eq:H1d_with_field}.
Finally, Fig.~\ref{fig:1dchain_speckle_page}(b) does not show obviously located outlying states when compared to Fig.~\ref{fig:1dchain_speckle_page}(a) as had been remarked earlier while discussing Fig.~\ref{fig:1dchain_level_stats}(a)~\cite{scar_entanglement_note}.

It must be noted that volume law entanglement is not forbidden in low-lying eigenstates \emph{including} the ground state~\cite{Gottesman_Hastings_2010,Irani_2010,Vitagliano_Riera_Latorre_2010,Ramirez_Sierra_2014,Bravyi_etal_2012,Movassagh_Shor_2016,Zhang_Ahmadain_Klich_2017,Salberger_Korepin_2018}. 
A recent Feynman-Kitaev clock based construction~\cite{Ippoliti_Long_2026} that avoids the specific ``rainbow'' nature of the previous constructions~\cite{Alexander_etal_2019,Alexander_etal_2021} takes it even further by engineering infinite temperature Floquet-like thermal behavior in all eigenstates including the ground state.
In light of this, our results are distinguished by being neither in the infinite temperature limit -- and in fact quite far from this limit near the spectral edge and rather in the zero temperature limit -- nor obeying area law of generic local Hamiltonians for low-lying eigenstates.
As mentioned above, the ground state of Eq.~\ref{eq:H1d_with_field} obeys the area law~\cite{supp_info}.
Thus the volume law observed near the spectral edge (Fig.~\ref{fig:1dchain_speckle_page}(b) inset) is to be understood as a consequence of the large thermodynamic entropy~\cite{Deutsch_Li_Sharma_2013,Dymarsky_Lashkari_Liu_2018} enabled by the EMICQ structure keeping in mind the order of limits mentioned above.

To argue for the spectral edge ETH effect in $2d$ in absence of a detailed finite size scaling study, let us look at the situation when one of the couplings dominates the other.
In such a situation, the many-body spectrum splits up into bands or lobes as may be seen in the extreme limits (e.g. $J^x \rightarrow 0$).
This is shown through the lobed structure of the many-body density of states in Fig.~\ref{fig:2dsquare_level_stats}(b).
Eq.~\ref{eq:2dchecker} has non-integrability ``within each lobe'' for \emph{all} lobes.
This is shown in Fig.~\ref{fig:2dsquare_level_stats}(c) by plotting the lobe-resolved average gap ratio by including only the eigenstates from the respective lobe.
As we can see the lobe-resolved gap ratios also tend to the Wigner-Dyson value.

In this lobe limit, the number of lobes is proportional to the number of unit cells, i.e. linear in the system size.
Every lobe has an overall exponential degeneracy due to the EMICQ structure.
In presence of a generic tiny perturbation such as an uniform field similar to the $1d$ case, all these states from different sectors will get coupled.
This will lead to the formation of a band of many-body states corresponding to each lobe due to the level repulsion brought in by the generic perturbation similar to the $1d$ case.
Thus each lobe will now become an extensive system of eigenstates in itself down to the lowest  lobe.
Furthermore, the lower excited lobes get pushed closer and closer to the spectral edge due the aforementioned linear scaling of lobes with system size.
This ensures the needed extensivity in terms of the many-body DOS or thermodynamic entropy for eigenstate thermalization. 
This extensivity property is generically ensured by an extensively large excitation energy, i.e. a finite $\epsilon$. 
For generic systems, this takes us into the high energy bulk.
For the non-integrable EMICQ model of Eq.~\ref{eq:2dchecker}, following the above argument, we have to inject only a small amount of excitation energy ($\epsilon \ll O(J)$) in the thermodynamic limit ($L \rightarrow \infty)$ to access an extensive thermodynamic entropy which make ETH operational near the spectral edge.
For the above argument, we keep in mind the order of limits discussion from earlier, i.e. taking the thermodynamic limit first and then the ``spectral edge'' limit.
This argument also works away from the lobe limit closer to $J^x \sim J^z$.
Finally, for the restricted $2d$ system sizes available, we show in Ref.~\cite{supp_info} that local observable matrix element and bipartite EE distributions are in line with the $1d$ results.

In conclusion, the EMICQ structure is a symmetry-based mechanism for engineering thermal behavior in lattice systems with infinitesimal injection of excitation energy. 
It would be exciting and potentially feasible to demonstrate the spectral edge ETH effect with present day capabilities such as in existing artificial quantum platforms.
This effect thus holds promise to engineer efficient entropy sinks at  ultra-low temperatures in quantum technologies.

Our results also provide the following perspective on thermalization in lattice gauge theories (LGTs) that has been the subject of recent attention~\cite{LGT_thermalization_review_2025}.
In our models, the local fluxes are conserved under the dynamics which is generically not the case in LGTs.
However, if they were to be approximately conserved, i.e. $[\hat{O},H] \sim \alpha \hbar \hat{O'}$ with the (dimensionless) coefficient $\alpha \ll 1$ for the local fluxes of some LGT, then the spectral edge ETH effect would be expected similar to the $1d$ results earlier.
Note that the mutual non-commutation or incompatibility of local fluxes is built in for non-Abelian LGTs.
It may thus be worthwhile to review the existing results on ETH physics in LGTs (e.g. see the recent Ref.~\cite{Chen_etal_su2lgt_thermalization_2026} for a summary) from this perspective provided by our results.

\prlsec{Acknowledgments} 
We acknowledge computational resources of the PARAM Rudra, Praganak and Chandra HPC clusters. 
S.~P. thanks Ajit Balram for a discussion.
A.~P. thanks Tanay Pathak and Abhik Kumar Saha for valuable discussions. 
S.~P. acknowledges support from ANRF-DST (formerly SERB), Govt. of India via Grant No. MTR/2022/000386.
M.~T. gratefully acknowledges support from JST CREST (Grant No. JPMJCR24I2) and from JSPS KAKENHI Grant Number JP25K00925.
A.~P., M.~T., and S.~P. were partially supported from the JST LOTUS Programme (Grant No. Z2025L8290004).

\bibliographystyle{apsrev4-2}
\bibliography{refs}

\end{document}


\title{Supplemental Material for ``Eigenstate Thermalization at the edge of many-body spectrum''}

\maketitle
\beginsupplement

In this document, we provide additional information in relation to the results presented in the main text for the EMICQ (extensively many mutually incompatible conserved quantities) models (Eq.~1-3 of the main text).

\section{1D Model: Level Statistics}
\label{suppsec:1d_level_statistics}
\begin{figure}[t]
    \centering
    \includegraphics[width=0.9\linewidth]{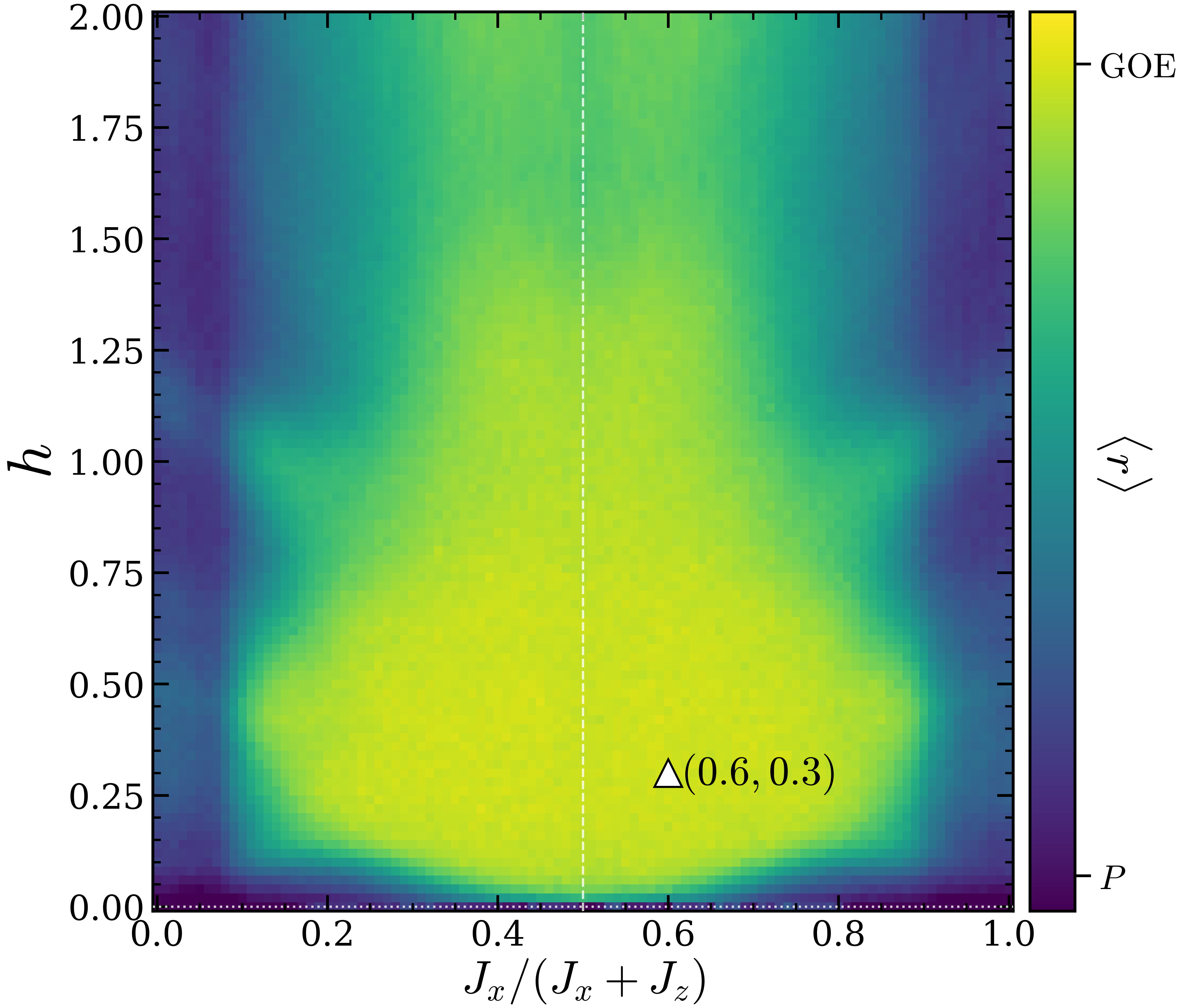}
    \caption{
        Mean adjacent-gap ratio $\langle r\rangle$ for the one-dimensional EMICQ chain model at $L=12$ as a function of the coupling ratio $J_x/(J_x+J_z)$ and the canted-field strength $h$.
        The Poisson and GOE values are $\langle r\rangle_{\mathrm{P}}\simeq0.3863$ and $\langle r\rangle_{\mathrm{GOE}}\simeq0.5307$, respectively.
        The triangle marks the parameter point $(J_x,J_z,h)=(0.6,0.4,0.3)$ used in the main text.
    }
    \label{suppfig:rmap_full}
\end{figure}

We first examine the level statistics of the one-dimensional EMICQ chain model for $L=12$ in the presence of the canted field (Eq.~3 of the main text).
For a sequence of ordered eigenenergies $E_n$, we define the adjacent level spacings as $\delta_n=E_{n+1}-E_n$, and the adjacent-gap ratio as $r_n=\frac{\min(\delta_n,\delta_{n+1})}{\max(\delta_n,\delta_{n+1})}$.
The adjacent-gap ratio is particularly useful since, unlike the conventional level-spacing distribution, it does not require an unfolding of the spectrum~\cite{OganesyanHuse2007}.
For an integrable spectrum with Poisson statistics, $\langle r\rangle_{\mathrm{P}}\simeq0.3863$, while a 
chaotic spectrum described by the Gaussian orthogonal ensemble (GOE) has $\langle r\rangle_{\mathrm{GOE}}\simeq0.5307$.
Figure~\ref{suppfig:rmap_full} shows the mean adjacent-gap ratio over a broad region of the parameter space spanned by the coupling ratio $J_x/(J_x+J_z)$ and the canted-field strength $h$.
Along the zero-field line, where the model is integrable, the level statistics remain close to the Poisson value.
Upon turning on the canted field, a broad region with $\langle r\rangle$ close to the GOE value develops for $L=12$.
Deviations from the GOE value are seen near the strongly anisotropic limits of the model and at large field strengths for the finite system size of $L=12$ considered here.
We are interested in parameter points with GOE level statistics since that ensures non-integrability.
The parameter point used for the main text analyses, $(J_x,J_z,h)=(0.6,0.4,0.3)$, is marked by the triangle in Fig.~\ref{suppfig:rmap_full}.

\begin{figure}[t]
    \centering

    \includegraphics[width=0.95\linewidth]{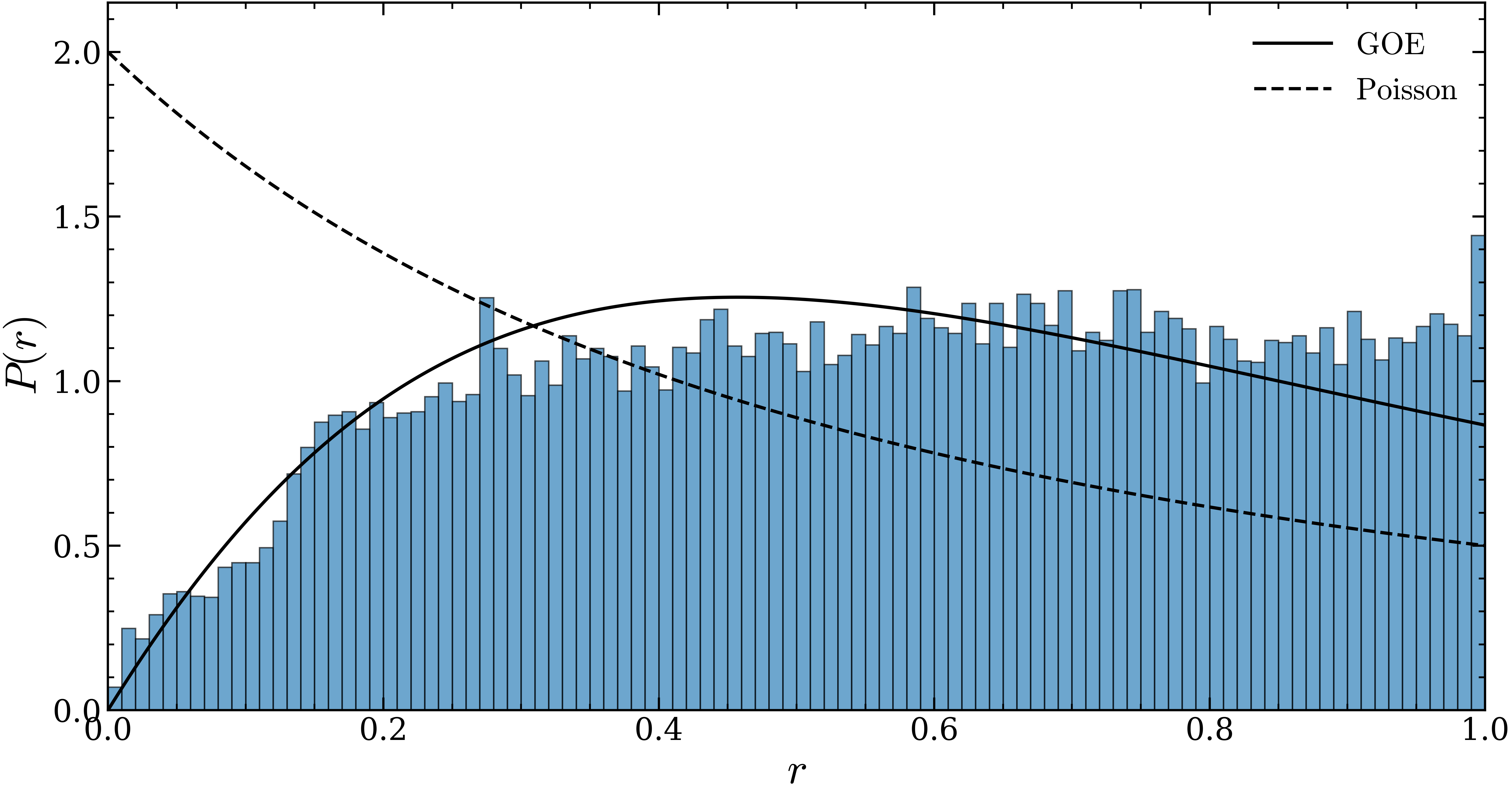}
    \caption{
    Distribution of the adjacent-gap ratio $P(r)$ for $(J_x,J_z,h)=(0.6,0.4,0.3)$ at $L=18$ in the $k=0$ sector.
    The solid and dashed curves show the GOE and Poisson predictions, respectively.
    The numerical distribution shows clear suppression of small gap ratios, indicating level repulsion, and is qualitatively consistent with GOE statistics rather than Poisson statistics.
    Finite-size deviations from the GOE prediction remain across the distribution, particularly at the smallest values of $r$ and toward $r\simeq1$.
}
    \label{suppfig:Rdist_GOE}
\end{figure}

To examine the level statistics more closely, we show the full adjacent-gap-ratio distribution $P(r)$ at the parameter point used in the main text.
For the definition of $r$ above, the Poisson and GOE distributions that we use as reference are~\cite{Atas2013}
\begin{equation}
 P_{\mathrm{P}}(r)=\frac{2}{(1+r)^2}
\end{equation}
and
\begin{equation}
    P_{\mathrm{GOE}}(r)=\frac{27}{4}\frac{r+r^2}{(1+r+r^2)^{5/2}}.
\end{equation}
The observed distribution is shown in Fig~\ref{suppfig:Rdist_GOE}. 
It exhibits clear level repulsion and is qualitatively much closer to the GOE expectation than to Poisson statistics.
Some deviations from the GOE distribution are seen that may be due to finite sizes, although the mean adjacent-gap ratio is observed to be very close to the GOE value as seen in Fig.~1 of the main text.

\section{1D Model: Local-operator eigenstate expectation values (EEV) and their fluctuations}
\label{suppsec:1d_local_operators}
\begin{figure*}[t]
    \centering

    \includegraphics[width=0.9\linewidth]{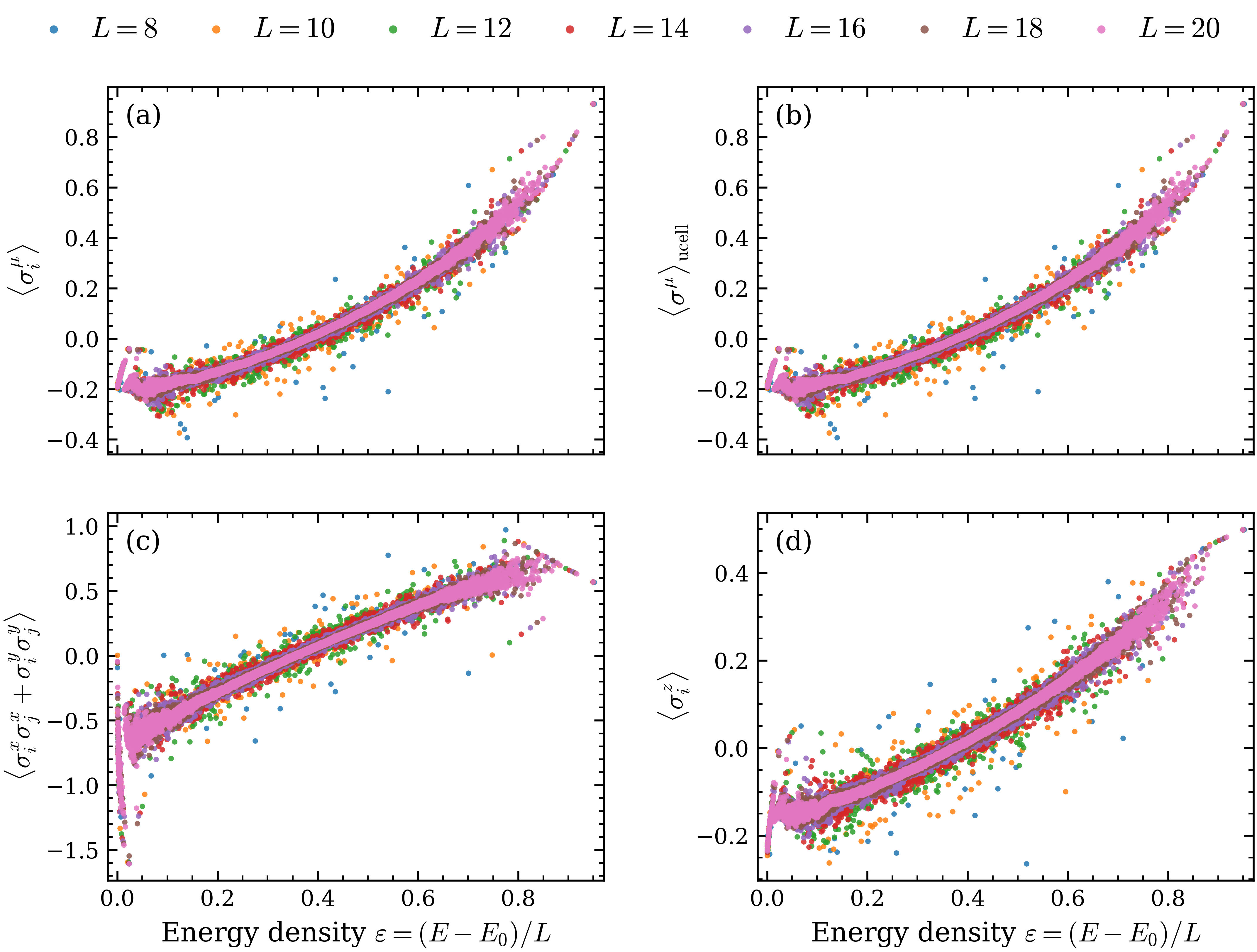}

    \caption{
        Eigenstate expectation values of representative local observables as functions of the excitation-energy density
        $\epsilon=(E-E_0)/L$ for
        $(J_x,J_z,h)=(0.6,0.4,0.3)$ in the $k=0$ sector.
        The panels show representative one-site and two-site observables defined in the text.
        For increasing system size, the EEV distributions become progressively narrower around their smooth energy-dependent backgrounds.
        The finite-size scaling of these fluctuations is analyzed in Fig.~\ref{suppfig:eev_fluctuation_scaling_raw_clipped}.
    }

    \label{suppfig:local_operator_eevs}
\end{figure*}
In the context of the analysis presented in Fig.~2a of the main text, We now examine the eigenstate expectation values (EEVs) of a larger set of local observables at the representative non-integrable point used in the main text, $(J_x,J_z,h)=(0.6,0.4,0.3)$.
For an eigenstate $|n\rangle$ with energy $E_n$, the diagonal matrix element of a local operator $\hat O$ is $O_{nn}=\langle n|\hat O|n\rangle$.
According to the eigenstate thermalization hypothesis, the diagonal matrix elements may be written as
\begin{equation}
    O_{nn}=\overline{O}(E_n)+e^{-S(E_n)/2}f_O(E_n,0)R_{nn}    
\end{equation}
where $\overline{O}(E)$ is a smooth function of energy, $S(E)$ is the thermodynamic entropy, and $R_{nn}$ is a fluctuating quantity with zero mean and variance of order unity.
Within a sufficiently narrow energy window, the typical eigenstate-to-eigenstate fluctuations $\sigma_O$ of $O_{nn}$ are therefore expected to scale as $\sigma_O\propto D_w^{-1/2}$, where $D_w$ is the number of eigenstates contained in that window~\cite{Beugeling_etal_pre_2014}.

We use the excitation-energy density $\epsilon=\frac{E-E_0}{L}$, where $E_0$ is the ground-state energy for the corresponding system size.
In addition to the local operator shown in the main text, we consider several representative one-site and two-site observables, $O_1=\langle\sigma_i^z\rangle$,$O_2=\langle\sigma_i^\mu\rangle$, $O_3=\langle\sigma_{\mathrm{ucell}}^\mu\rangle$,$O_4=\left\langle\sigma_i^x\sigma_j^x+\sigma_i^y\sigma_j^y\right\rangle$, and $ O_5=\langle\sigma_i^\mu\sigma_j^\mu\rangle$.
Here $\mu$ denotes the same generic direction in spin space used for the corresponding local operators.
Figure~\ref{suppfig:local_operator_eevs} shows the EEVs of these observables as a function of excitation-energy density.
For all of the operators, the eigenstate expectation values form a smooth energy-dependent background together with finite-size fluctuations about that background.
The distribution becomes progressively narrower with increasing system size, both in the spectral bulk and close to the lower spectral edge.

\begin{figure*}[t]
    \centering
    \includegraphics[width=0.95\linewidth]{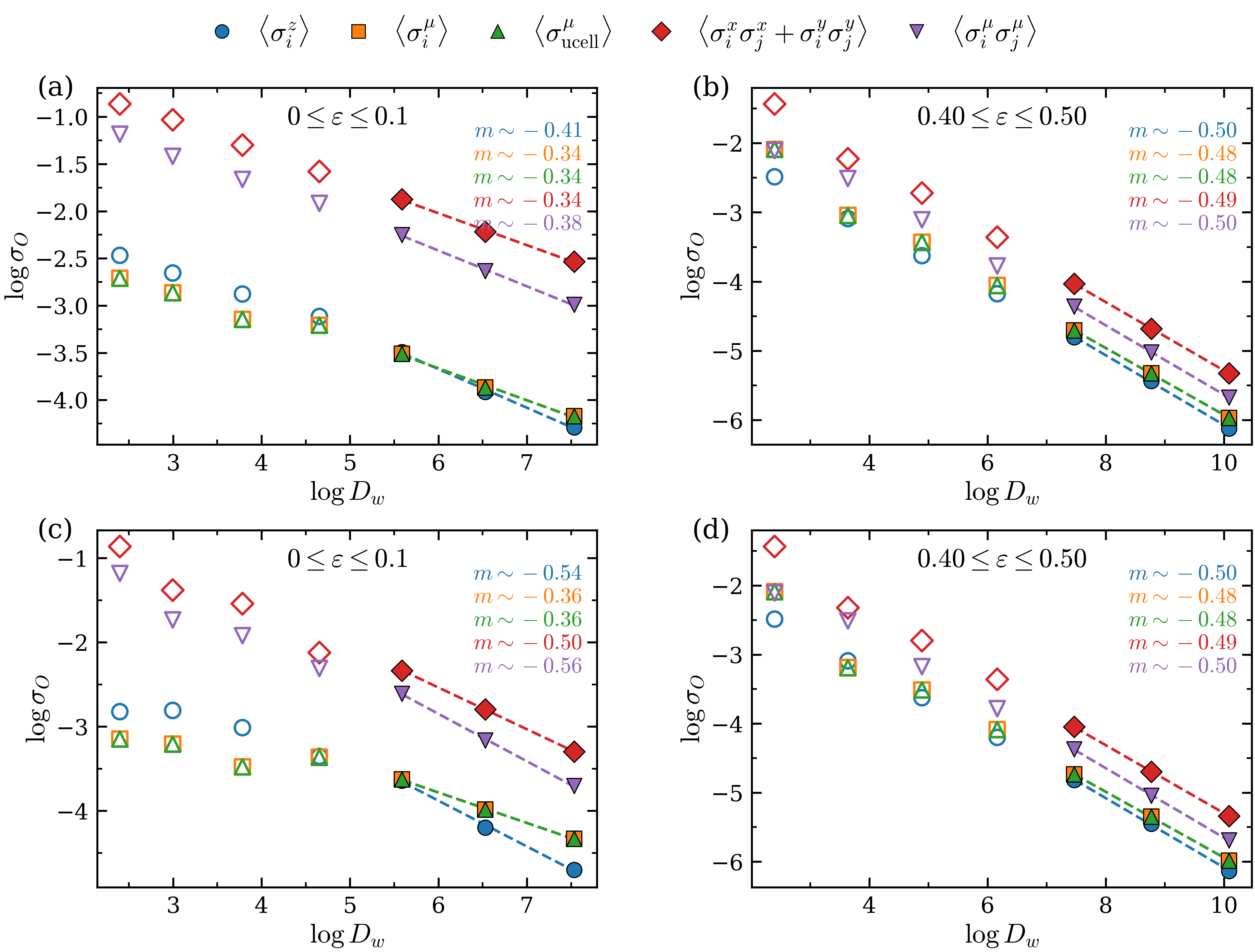}
    \caption{
        Finite-size scaling of local-observable EEV fluctuations with the Hilbert-space window dimension $D_w$ for $(J_x,J_z,h)=(0.6,0.4,0.3)$ in the $k=0$ sector.
        Panels (a) and (b) show the raw fluctuations in the lower-edge window $0\leq\epsilon\leq0.1$ and the bulk window $0.40\leq\epsilon\leq0.50$, respectively.
        Panels (c) and (d) show the corresponding results after removing the large-deviation states according to the clipping procedure described in Sec.~\ref{subsec:outlier_clipping}.
        Filled markers denote the system sizes included in the power-law fits, while open markers denote the smaller sizes excluded from the fits.
        Dashed lines show fits of the form $\sigma_O\propto D_w^m$, where $m$ denotes the fitted scaling exponent.
        In the spectral bulk, both the unclipped and clipped distributions exhibit exponents very close to the ETH expectation $m=-1/2$.
        Near the spectral edge, the unclipped scaling is systematically slower, while clipping brings several of the observables closer to the ETH expectation, although some observable dependence remains.
    }

    \label{suppfig:eev_fluctuation_scaling_raw_clipped}
\end{figure*}

We characterize the finite-size scaling through the exponent $m$ defined by $\sigma_O\propto D_w^m$.
To quantify the finite-size suppression of these fluctuations, we consider two fixed excitation-energy density windows.
We take the window $\epsilon\in[0.00,0.10]$ near the lower edge of the spectrum, while the window $\epsilon\in[0.40,0.50]$ representative of the bulk part of the spectrum.
For each observable and each system size, we first remove the smooth energy dependence within the corresponding energy window and calculate the standard deviation $\sigma_O$ of the resulting residual fluctuations.
Panels (a) and (b) of Fig.~\ref{suppfig:eev_fluctuation_scaling_raw_clipped} show the resulting raw fluctuations as a function of the number of states $D_w$ in the window for the edge and bulk, respectively.
Near the spectral edge, the fluctuations decrease with increasing $D_w$, with fitted exponents ranging approximately from $-0.33$ to $-0.41$ for the different local observables.
The bulk fluctuations whereas exhibit scaling very close to the ETH expectation, with fitted exponents between approximately $-0.49$ and $-0.50$.


\subsection{Outlier states and clipping procedure}
\label{subsec:outlier_clipping}
The EEV distributions in Fig.~\ref{suppfig:local_operator_eevs} show a (small) number of states lying well outside the dense cluster formed by the majority of eigenstates.
To identify these states, we first construct a smooth ``microcanonical'' background $\overline{O}(E)$ for each observable and system size.
The spectrum is divided into equal energy bins, and the median EEV within each bin is calculated.
A further smoothening step is carried out by computing the median of the binned medians from a set of contiguous bins (five contiguous bins was used).
A smooth interpolation is constructed out of the median-of-binned-medians data set to arrive at $\overline{O}(E)$.
The residual for each eigenstate is then defined as $\delta O_n=O_{nn}-\overline{O}(E_n)$.
In each energy bin, we estimate the width of the residual distribution using the median of the (absolute value) of the residuals as
\begin{equation}
    \sigma_{\mathrm{MAD}}=1.4826\,\mathrm{median}\left(|\delta O_n-\mathrm{median}(\{\delta O_n\})|\right),
\end{equation}
where $\{\delta O_n\}$ is set of residuals for a given energy bin.
The factor $1.4826$ gives the Gaussian-equivalent standard deviation.
An eigenstate is classified as an outlier when 
\begin{equation}
    |\delta O_n-\mathrm{median}(\{\delta O_n\})| > 3\sigma_{\mathrm{MAD}}
    \label{suppeq:3sigma_criterion}
\end{equation}
This criterion prevents the large-deviation states themselves from strongly affecting the scale used to identify them which is desirable.
The states satisfying the above outlier criterion are excluded and the EEV fluctuations are recalculated.
Panels (c) and (d) of Fig.~\ref{suppfig:eev_fluctuation_scaling_raw_clipped} show the resulting scaling in the lower-edge and bulk windows for the clipped EEVs, respectively.
For the lower-edge window, the fitted exponents now become close to $-0.50$ for several observables unlike in panel (a) of Fig.~\ref{suppfig:eev_fluctuation_scaling_raw_clipped}.
In the bulk window, the fitted exponents remain close to $-0.5$ as in panel (b) of Fig.~\ref{suppfig:eev_fluctuation_scaling_raw_clipped}.


\subsection{Outlier fraction}
\label{subsec:outlier_fraction}
\begin{figure}[t]
    \centering

    \includegraphics[width=0.9\linewidth]{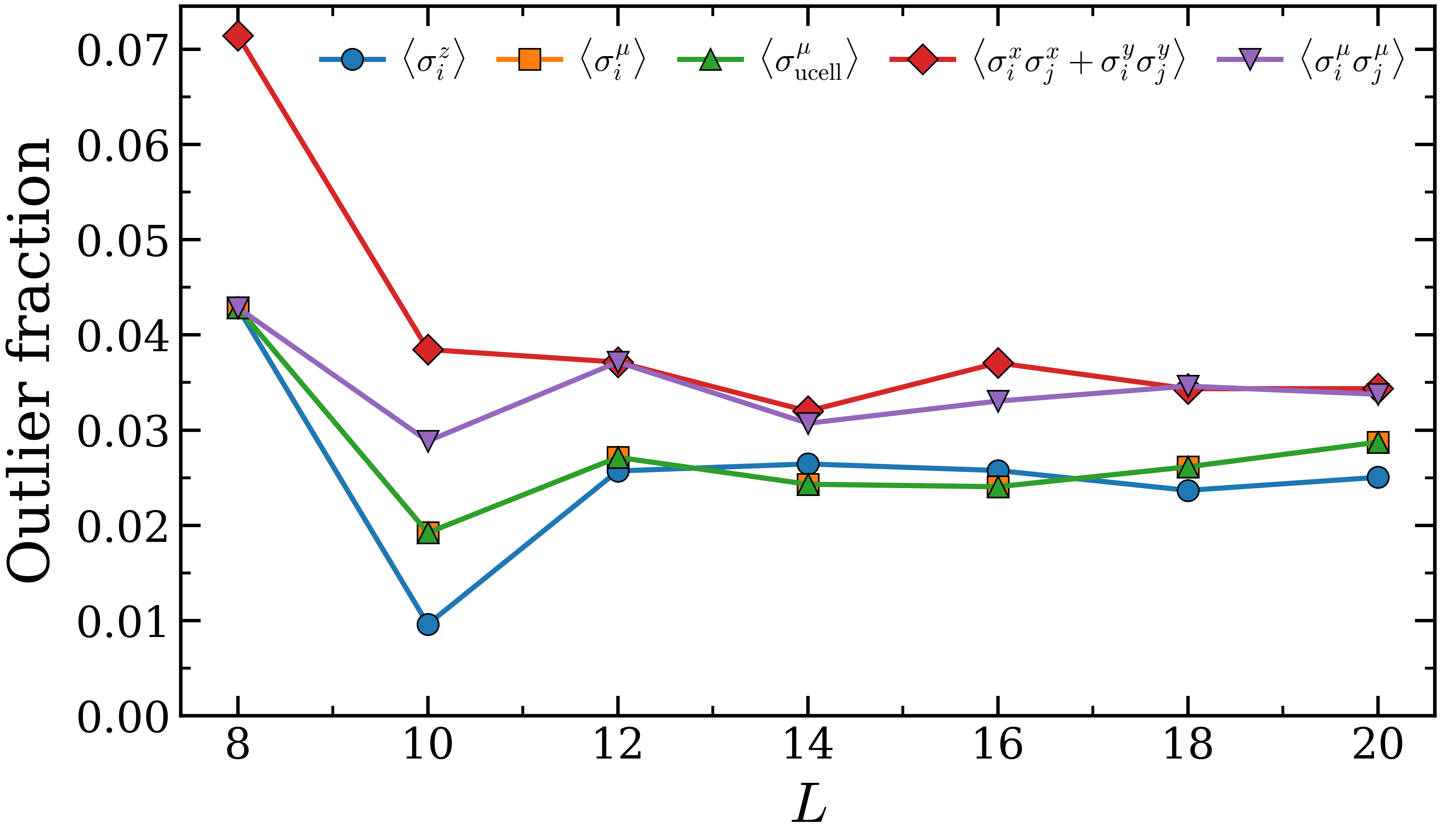}

    \caption{
        Full-spectrum outlier fraction as a function of system size $L$ for the five local observables at $(J_x,J_z,h)=(0.6,0.4,0.3)$ in the $k=0$ sector.
        An eigenstate is classified as an outlier when its residual satisfies $|\delta O_n-\mathrm{median}(\delta O)|>3\sigma_{\mathrm{MAD}}$.
        The outlier fraction remains at the few-percent level for the larger system sizes, substantially above the approximately $0.27\%$ weight expected outside $\pm3\sigma$ for a Gaussian distribution.
    }

    \label{suppfig:outlier_fraction}
\end{figure}

We also track how frequently the large-deviation states occur as the system size is increased.
For each observable, we define the outlier fraction as
\begin{equation}
    f_{\mathrm{out}}=\frac{N_{\mathrm{out}}}{N_{\mathrm{tot}}},
\end{equation}
where $N_{\mathrm{out}}$ is the number of eigenstates satisfying the $3\sigma_{\mathrm{MAD}}$ criterion of Eq.~\ref{suppeq:3sigma_criterion} and $N_{\mathrm{tot}}$ is the total number of eigenstates (within the symmetry sector).
Figure~\ref{suppfig:outlier_fraction} shows $f_{\mathrm{out}}$ as a function of system size for the five local observables.
Apart from stronger finite-size variations at the smallest sizes, the outlier fraction remains at the level of a few percent over the accessible range of $L$.
The observed fractions are larger than would be expected from a purely Gaussian distribution of the EEVs within an energy bin.
At the same time, the outliers remain a small fraction of the complete spectrum.

\section{1D Model: Entanglement Entropy}
\label{suppsec:1d_entanglement}

\subsection{Dependence on the excitation-energy window width}
\label{subsec:ee_window_width}
\begin{figure}[t]
    \centering

    \includegraphics[width=0.95\linewidth]{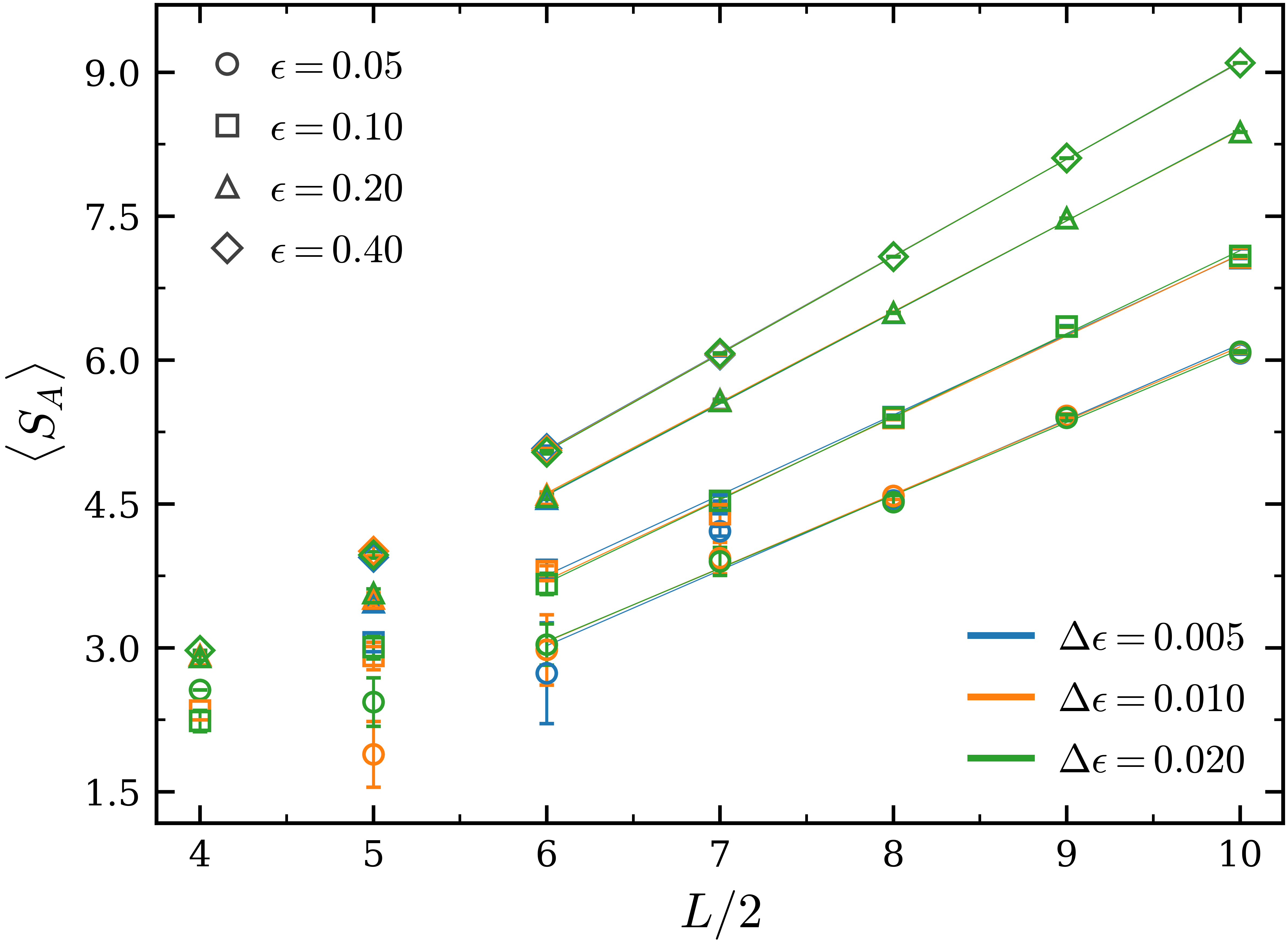}
    \caption{
        Dependence of the mean half-chain entanglement entropy $\langle S_A\rangle$ on the excitation-energy window width.
        Different marker shapes denote windows centred at
        $\epsilon=0.05$, $0.10$, $0.20$, and $0.40$, while the colors denote
        $\Delta\epsilon=0.005$, $0.01$, and $0.02$.
        For each fixed excitation-energy density, the results obtained with the three window widths closely track one another, particularly at the larger system sizes.
        The volume-law growth of $\langle S_A\rangle$ with $L/2$ is therefore insensitive to the precise choice of $\Delta\epsilon$ over the range considered.
    }

    \label{suppfig:ee_window_width}
\end{figure}

In the context of the entanglement data analysis presented in Fig.~2b of the main text, we now check that the volume-law scaling is not sensitive to the chosen width of the excitation-energy window, we repeat the analysis for several values of $\Delta\epsilon$.
Figure~\ref{suppfig:ee_window_width} shows the mean half-chain entanglement entropy $\langle S_A\rangle$ as a function of subsystem size $L/2$ for windows centered at
$\epsilon=0.05$, $0.10$, $0.20$, and $0.40$.
For each fixed excitation-energy density, we compare window widths $\Delta\epsilon=0.005,\;0.01,\;\text{and}\;0.02.$
The resulting curves are nearly indistinguishable over most of the accessible system sizes.
The dependence on $\Delta\epsilon$ is visible for the smallest system sizes and for the lowest excitation-energy windows, where only a small number of eigenstates contribute to the average.
These differences rapidly decrease with increasing system size.
The linear growth of $\langle S_A\rangle$ with $L/2$ is thus robust to the choice of the width of the energy window.

\subsection{Momentum-sector dependence}
\label{subsec:ee_momentum_sectors}
\begin{figure}[t]
    \centering
    \includegraphics[width=0.95\linewidth]{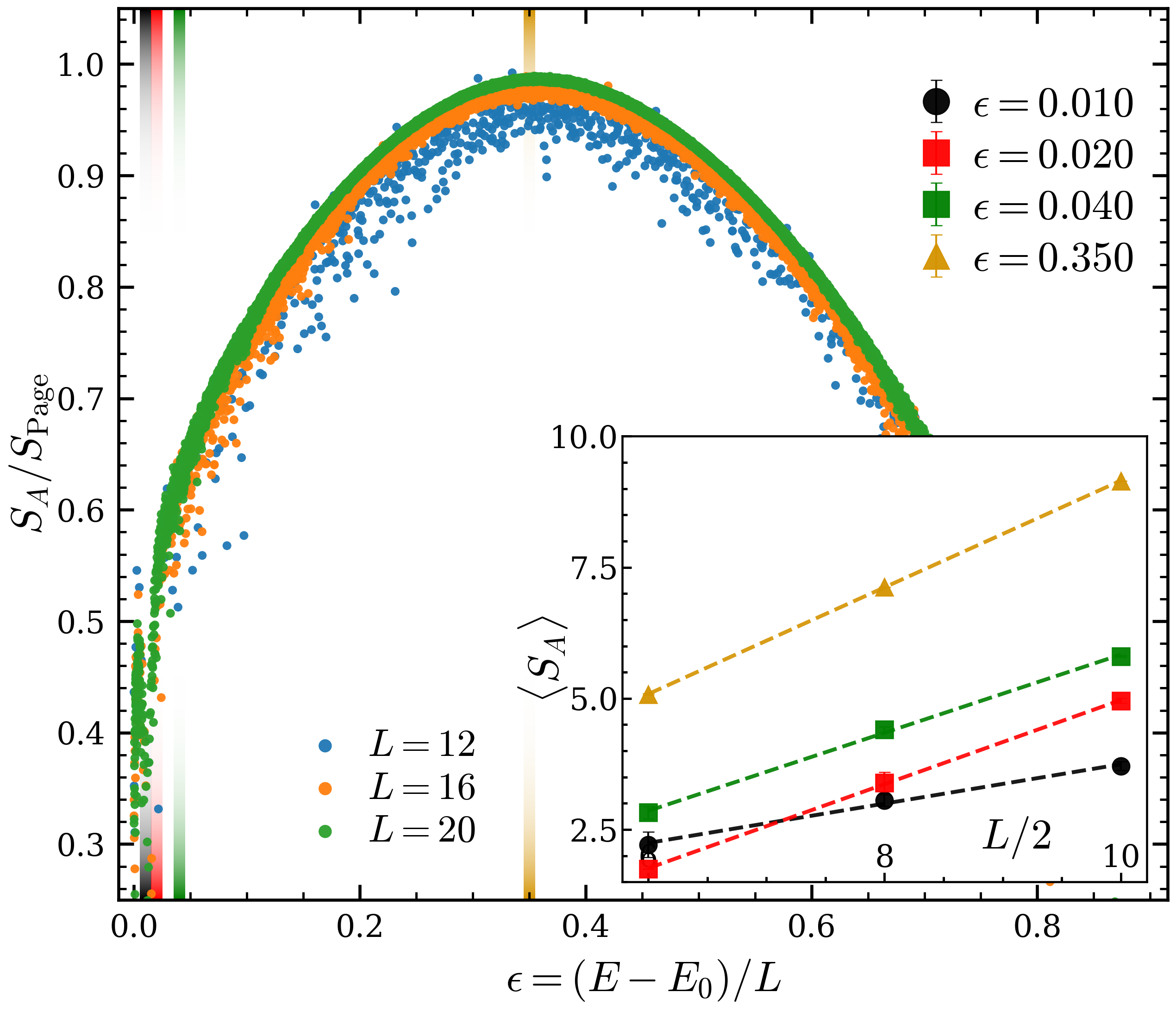}
    \caption{
    Half-chain entanglement entropy in the $k=\pi$ momentum sector at the same non-integrable point as above.
    The main panel shows the Page-normalized entanglement entropy as a function of the excitation energy density $\epsilon$ for $L=12, 16$ and $20$.
    The inset shows the mean entanglement entropy for a window of width $\Delta\epsilon=0.01$ centered at $\epsilon=0.01, 0.02, 0.04$ and $0.35$ as a function of $L/2$.
    }
    \label{suppfig:ee_kpi}
\end{figure}

We also check the entanglement behavior discussed above for a different momentum sector.
Figure~\ref{suppfig:ee_kpi} shows the Page-normalized half-chain entanglement entropy in the $k=\pi$ sector at the same non-integrable parameter point $(J_x,J_z,h)=(0.6,0.4,0.3)$.
The inset shows the mean entanglement entropy $\langle S_A\rangle$ for several fixed nonzero excitation-energy densities.
The results are very similar to that in the $k=0$ sector discussed in Fig. 2b of the main text.

\subsection{Ground-state entanglement}
\label{subsec:ee_ground_state}

\begin{figure}[t]
    \centering
    \includegraphics[width=0.95\linewidth]{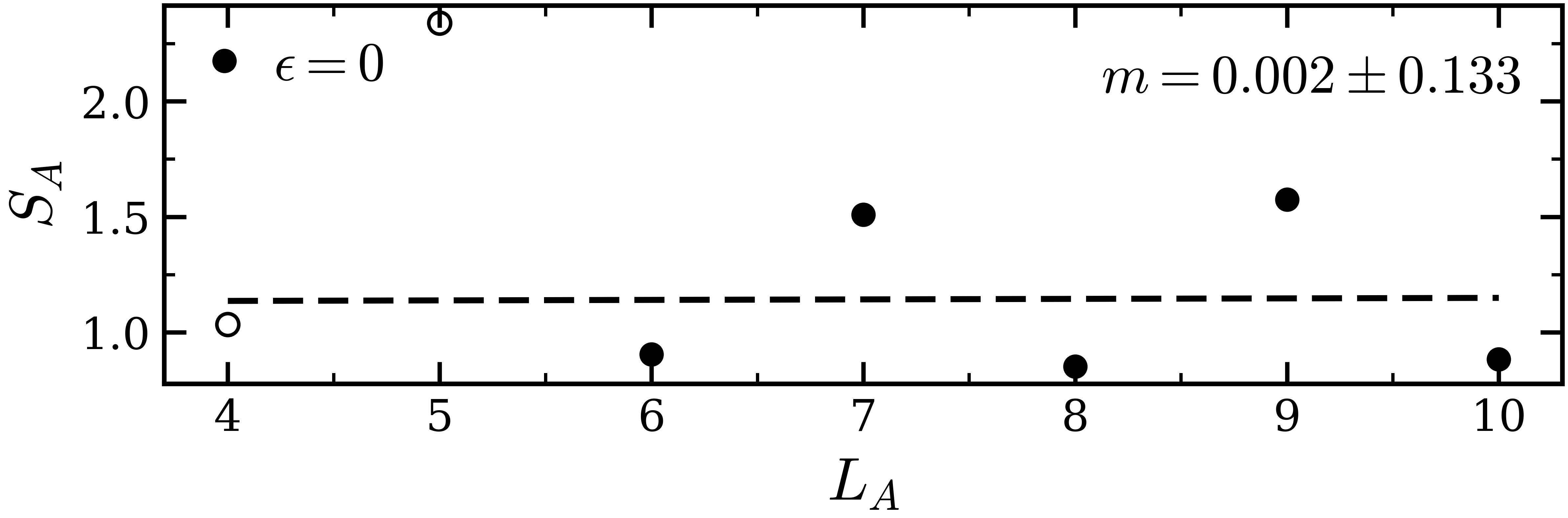}
    \caption{
        Entanglement entropy $S_A$ of the lowest-energy state as a function of subsystem size $L_A$ for the parameter point $(J_x,J_z,h)=(0.6,0.4,0.3)$ in the $k=0$ sector.
        Open markers denote the smaller system sizes excluded from the fit, while filled markers denote the sizes included in the fit.
        The dashed line shows a linear fit of the form $S_A=mL_A+b$, with the fitted slope indicated in the panel.
        Within the fitting uncertainty, the extracted slope is consistent with zero, indicating that the lowest-energy state does not exhibit volume-law entanglement over the accessible system sizes.
    }
    \label{suppfig:ground_state_ee}
\end{figure}

We now look at the entanglement of the ground state in the context of the order of limits discussion in the main text.
Figure~\ref{suppfig:ground_state_ee} shows the half-chain entanglement entropy $S_A$ of the ground state as a function of subsystem size $L_A=L/2$.
Unlike the finite-$\epsilon$ results discussed above, the ground-state entanglement does not show a systematic increase with subsystem size.
Instead, $S_A$ remains of order unity over the accessible system sizes and displays an even--odd finite-size oscillation.
Attempting a linear fit gives a slope $0.002\pm0.133$, which is consistent with zero within the fitting uncertainty.
The data of Fig.~\ref{suppfig:ground_state_ee} is therefore consistent with area-law entanglement in the ground state that is generally expected for local Hamiltonians.
This provides a direct contrast with the behavior at any fixed nonzero excitation-energy density considered above, where $\langle S_A\rangle$ grows linearly with $L_A$.
We understand these results based on the order-of-limits distinction discussed in the main text.
\begin{figure}[h]
    \centering

    \includegraphics[width=0.9\linewidth]{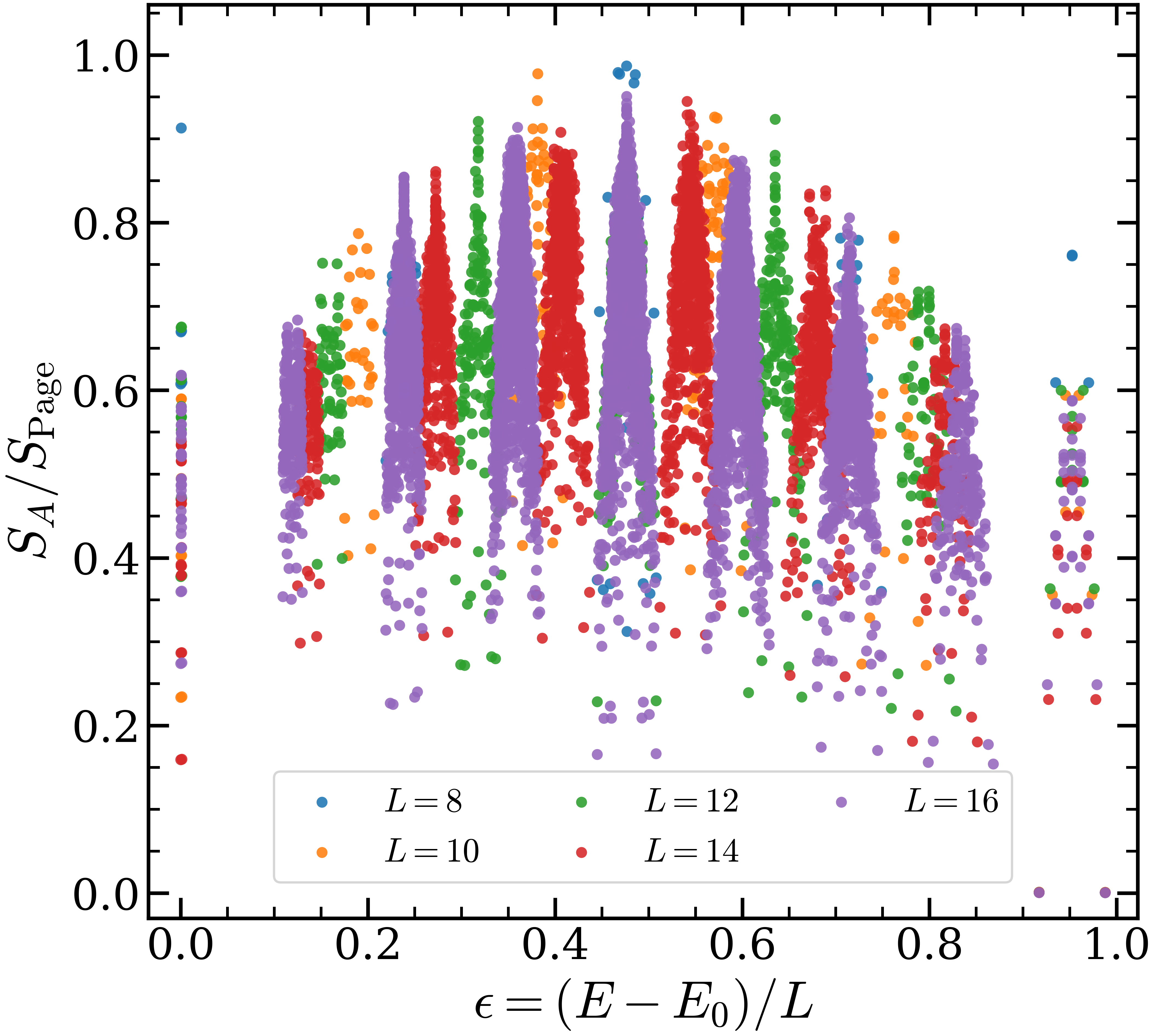}

    \caption{
        Half-chain entanglement entropy, normalized by the corresponding Page value, as a function of excitation-energy density
        $\epsilon=(E-E_0)/L$ at the integrable point
        $(J_x,J_z,h)=(0.05,0.95,0.05)$.
        In contrast to the non-integrable regime, the spectrum exhibits pronounced banding and broad eigenstate-to-eigenstate variations in the entanglement entropy.
        Low-entanglement states persist throughout a substantial part of the spectrum, and the data do not approach a single smooth Page-like envelope over the system sizes shown.
    }

    \label{suppfig:ee_integrable}
\end{figure}
 
\subsection{Entanglement entropy in the integrable limit}
\label{subsec:ee_integrable_limit}
For comparison, we also examine the entanglement structure at the parameter point $(J_x,J_z,h)=(0.05,0.95,0.05)$ which is very close to integrability with an average gap ratio around $\sim 0.381$.
Figure~\ref{suppfig:ee_integrable} shows the half-chain entanglement entropy, normalized by the corresponding Page value, as a function of excitation-energy density.
The structure is qualitatively different from that found in the non-integrable regime.
Rather than forming a smooth entanglement profile with progressively reduced eigenstate-to-eigenstate fluctuations, the integrable spectrum displays pronounced banding and a broad distribution of entanglement values throughout the spectrum.
States with substantially reduced entanglement persist even at intermediate excitation-energy densities, and the distributions do not collapse onto a single smooth Page-like envelope with increasing system size.
This provides a useful contrast with the non-integrable regime studied in the main text, where the entanglement entropy becomes increasingly smooth as a function of energy and exhibits volume-law scaling at fixed nonzero excitation-energy densities including near the spectral edge.


\section{2D Checkerboard Model}
\label{suppsec:2d_checkerboard}

\subsection{Level statistics}
\label{subsec:2d_level_statistics}
\begin{figure}[t]
    \centering

    \includegraphics[width=0.9\linewidth]{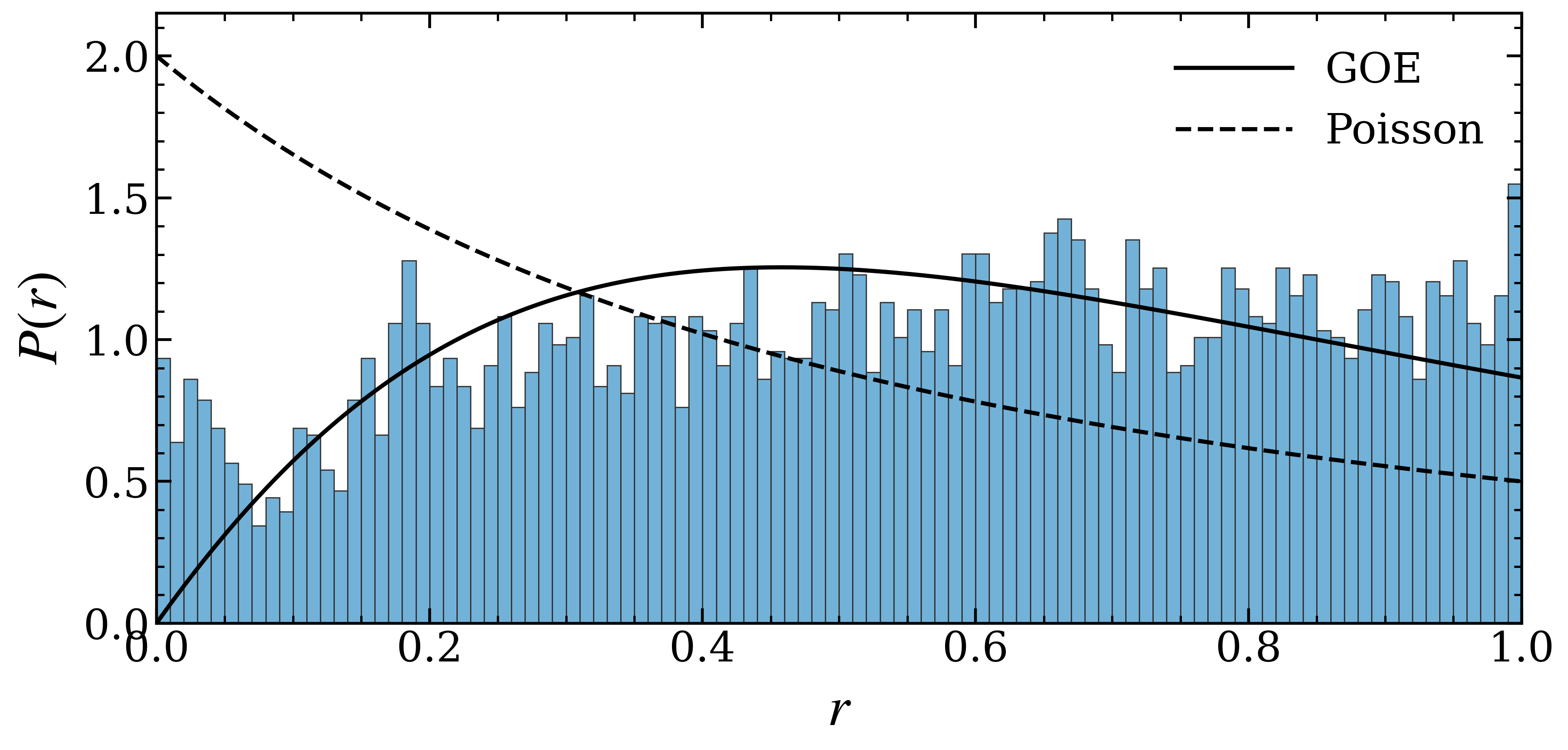}

    \caption{
        Distribution of the adjacent-gap ratio $P(r)$ for the $L=16$ checkerboard model at $J_x=J_z=0.5$ within a fixed anticommuting symmetry sector (the plaquette $\mathbb{Z}_2$ parities on $\boxed{x}$ plaquettes set to $+1$).
        The solid and dashed curves show the GOE and Poisson reference distributions, respectively~\cite{Atas2013}.
        The numerical distribution exhibits level repulsion and is qualitatively closer to GOE statistics than to Poisson statistics over much of the interval, while appreciable finite-size deviations remain, particularly at small $r$.
        The corresponding mean gap ratio is $\langle r\rangle=0.532$, very close to the GOE value $\langle r\rangle_{\mathrm{GOE}}\simeq0.5307$.
    }

    \label{suppfig:checkerboard_pr}
\end{figure}

We first examine the adjacent-gap-ratio distribution of the two-dimensional checkerboard EMICQ model (Eq.~1 of the main text).
Figure~\ref{suppfig:checkerboard_pr} shows $P(r)$ for the $L=16$ checkerboard cluster at the isotropic point $J_x=J_z=0.5$, within a fixed anticommuting symmetry sector (the plaquette $\mathbb{Z}_2$ parities on $\boxed{x}$ plaquettes set to $+1$).
The corresponding mean adjacent-gap ratio is $\langle r\rangle = 0.532$, which is very close to the GOE value $\langle r\rangle_{\mathrm{GOE}}\simeq0.5307$.
As in the one-dimensional analysis, we use the adjacent-gap ratio because it does not require an unfolding of the many-body spectrum.
The numerical distribution shows clear suppression at small $r$ indicating level repulsion.
Over a substantial part of the interval, the distribution is qualitatively closer to the GOE form than to the Poisson result, although noticeable finite-size deviations remain.
In particular, excess weight is visible at small $r$, showing that the finite-size spectrum is not fully captured by the ideal GOE distribution.

\subsection{Local-observable eigenstate expectation values}
\label{subsec:2d_local_observables}

\begin{figure}[t]
    \centering

    \includegraphics[width=0.95\linewidth]{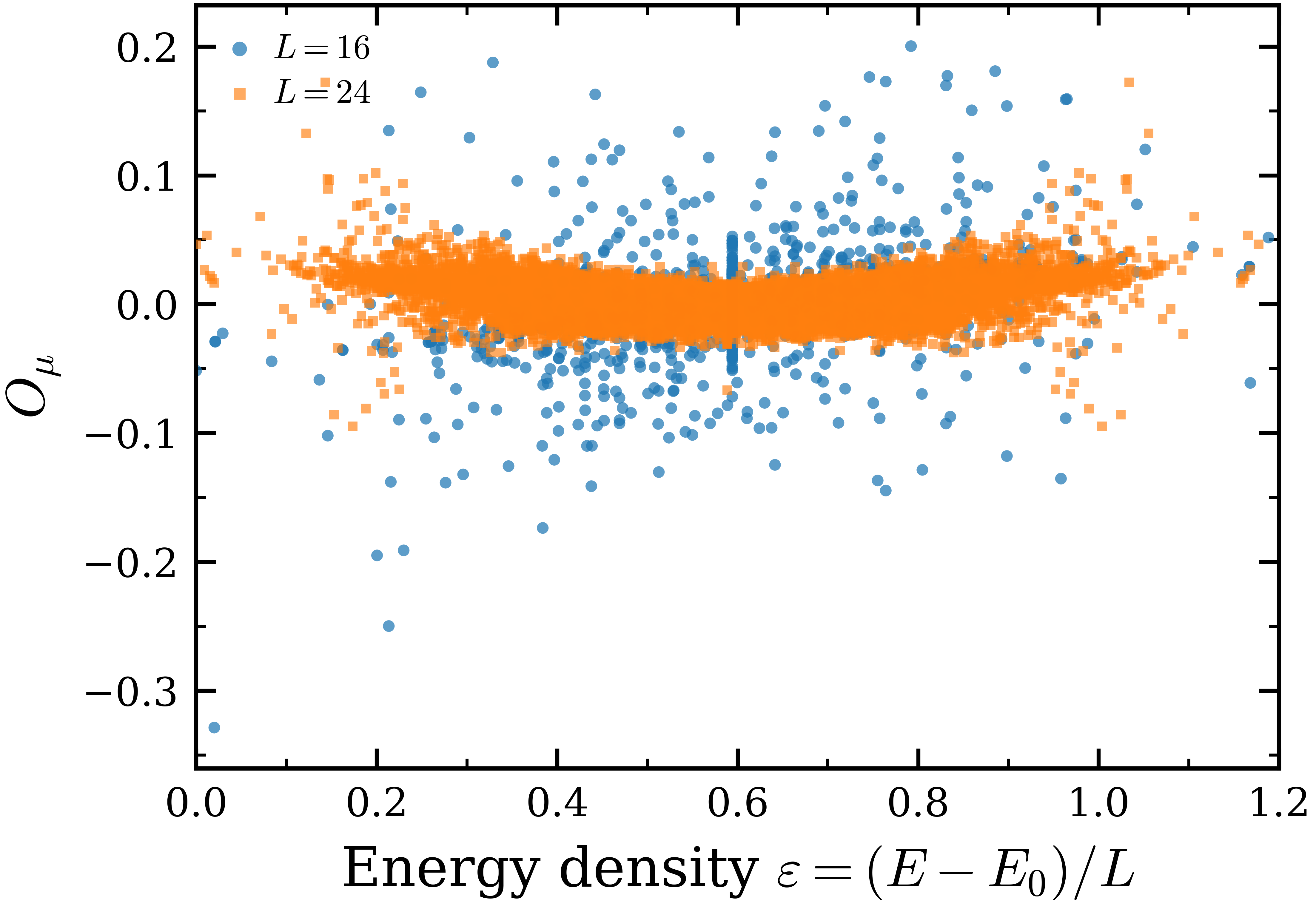}

    \caption{
        Eigenstate expectation values of the local bond observable $O_\mu$
        as a function of excitation-energy density
        $\epsilon=(E-E_0)/L$ for the $L=16$ and $L=24$ checkerboard clusters.
        Both datasets are obtained within a fixed anticommuting symmetry sector (the plaquette $\mathbb{Z}_2$ parities on $\boxed{x}$ plaquettes set to $+1$) and the $\mathbf{k}=(0,0)$ momentum sector.
        The EEV distribution becomes substantially narrower upon increasing the system size.
        In the lower-energy window, the fluctuation-width ratio is
        $\sigma_O^{(24)}/\sigma_O^{(16)}\simeq0.236$, compared with the ETH estimate $\sqrt{D_{16}/D_{24}}\simeq0.266$.
        In the spectral bulk, the corresponding values are approximately $0.153$ and $0.145$, respectively.
    }

    \label{suppfig:checkerboard_eev_comparison}
\end{figure}

We next examine the finite-size evolution of local-observable eigenstate expectation values in the checkerboard model.
Figure~\ref{suppfig:checkerboard_eev_comparison} compares the EEV distributions for the $L=16$ and $L=24$ clusters in the fixed anticommuting symmetry sector (the plaquette $\mathbb{Z}_2$ parities on $\boxed{x}$ plaquettes set to $+1$) and $\mathbf{k}=(0,0)$ momentum sector.
We consider the local bond observable $O_\mu=\langle \sigma_i^\mu \sigma_j^\mu\rangle$.
The EEVs are shown as a function of the excitation-energy density.
A clear narrowing of the distribution is visible upon increasing the system size from $L=16$ to $L=24$.
To quantify this suppression of eigenstate-to-eigenstate fluctuations, we compare the raw standard deviation $\sigma_O$ within fixed excitation-energy-density windows.
For the lower-energy window $\epsilon \in [0.0, 0.2]$, the number of states increases from $D_{16}=29$ to $D_{24}=409$.
The corresponding fluctuation widths are $\sigma_O^{(16)}=0.1012,\qquad \sigma_O^{(24)}=0.0239$, giving $\frac{\sigma_O^{(24)}}{\sigma_O^{(16)}}
\simeq0.236$.
For ETH-like fluctuations, the expected finite-size suppression is $\sqrt{\frac{D_{16}}{D_{24}}}\simeq0.266$.
The observed ratio is thus close to the ETH estimate.
A similar comparison can be made in the spectral bulk, using the window $0.35\leq\epsilon\leq0.55$.
Here the number of states grows from $D_{16}=319$ to $D_{24}=15128$, while the fluctuation widths decrease from $\sigma_O^{(16)}=0.0444$ to $\sigma_O^{(24)}=0.00679$.
This gives $\frac{\sigma_O^{(24)}}{\sigma_O^{(16)}}\simeq0.153$, again close to the ETH estimate $\sqrt{\frac{D_{16}}{D_{24}}}\simeq0.145$.
Thus, despite having access to only two system sizes in the checkerboard geometry, the strong narrowing of the EEV distribution from $L=16$ to $L=24$ is quantitatively consistent with the finite-size suppression expected for ETH-like fluctuations, both in the spectral bulk and towards the edge of the spectrum.

\subsection{Entanglement entropy}
\label{subsec:2d_entanglement}

\begin{figure}[t]
    \centering

    \includegraphics[width=0.95\linewidth]{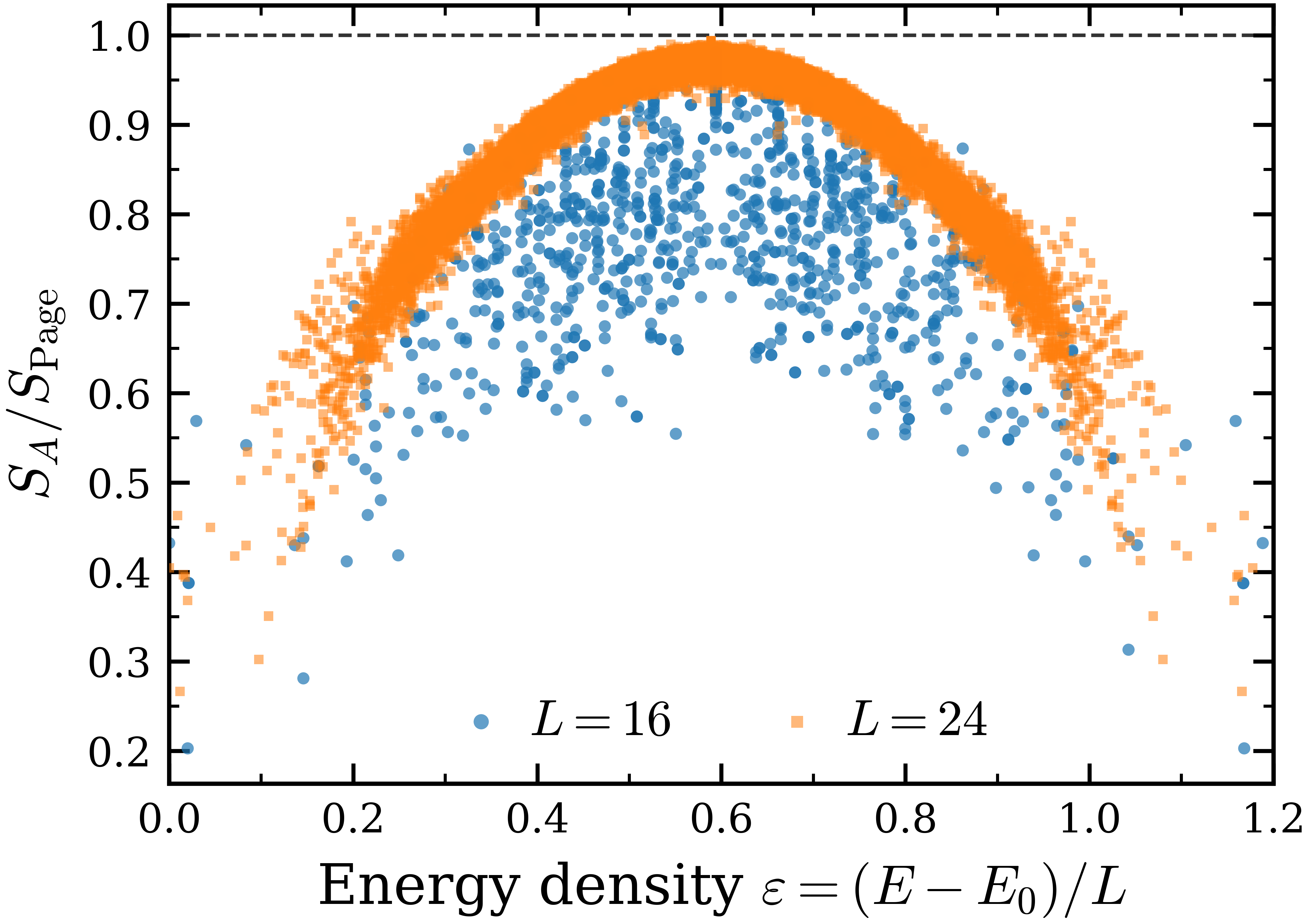}

    \caption{
        Half-system entanglement entropy normalized by the corresponding sector-resolved Page value as a function of excitation-energy density
        $\epsilon=(E-E_0)/L$ for the $L=16$ and $L=24$ checkerboard clusters.
        The calculations are performed in the fixed anticommuting symmetry sector (the plaquette $\mathbb{Z}_2$ parities on $\boxed{x}$ plaquettes set to $+1$), with $\mathbf{k}=(0,0)$ in both cases.
        For $L=16$, the parity-restricted half-system dimensions are $d_A=d_B=64$, giving $S_{\mathrm{Page}}\simeq5.279$ bits.
        For $L=24$, $d_A=d_B=512$, giving $S_{\mathrm{Page}}\simeq8.279$ bits.
        The $L=24$ distribution forms a substantially narrower envelope and approaches the Page value more closely in the high-entropy region of the spectrum.
    }

    \label{suppfig:checkerboard_ee_comparison}
\end{figure}

We finally compare the bipartite entanglement entropy of the $L=16$ and $L=24$ checkerboard clusters.
Figure~\ref{suppfig:checkerboard_ee_comparison} shows the half-system entanglement entropy as a function of excitation-energy density, normalized by the appropriate Page value for the symmetry sector.
For both $L=16$ and $L=24$, the calculation is performed in the fixed anticommuting symmetry sector (the plaquette $\mathbb{Z}_2$ parities on $\boxed{x}$ plaquettes set to $+1$) and the $\mathbf{k}=(0,0)$ momentum sector.
For the $L=16$ cluster, the half-system bipartition contains $L_A=L_B=8$ sites.
Within the anticommuting symmetry sector, the corresponding subsystem dimensions are $d_A=d_B=64$.
For a random pure state in a bipartite Hilbert space with $d_A\leq d_B$, the average Page entropy~\cite{Page1993} is
\begin{equation}
    S_{\mathrm{Page}}=\frac{1}{\ln 2}\left[\sum_{n=d_B+1}^{d_A d_B}\frac{1}{n}-\frac{d_A-1}{2d_B}\right],
\end{equation}
where the factor $1/\ln 2$ expresses the entropy in bits.
For $d_A=d_B=64$, this gives $S_{\mathrm{Page}}^{(16)}\simeq 5.279$, while the maximum possible entropy in the same reduced Hilbert space is $S_{\mathrm{max}}^{(16)}=\log_2 64=6$.
For the $L=24$ cluster, the half-system bipartition contains $L_A=L_B=12$ sites.
Within the anticommuting symmetry sector, the corresponding subsystem dimensions are $d_A=d_B=512$, giving $S_{\mathrm{Page}}^{(24)}\simeq 8.279$, with $S_{\mathrm{max}}^{(24)}=\log_2 512=9$.
The Page normalization above is defined from the anticommuting symmetry restricted bipartite Hilbert-space dimensions.
The comparison between the two sizes shows a pronounced finite-size narrowing of the entanglement distribution.
The reduction of the entanglement fluctuations with increasing system size is consistent with the finite-size evolution expected for typical thermal eigenstates.


\bibliography{refs}